\documentclass[twocolumn,secnumarabic,amssymb, nobibnotes, aps, prd, superscriptaddress]{revtex4-1}
\usepackage{graphicx}
\usepackage{dcolumn}
\usepackage{bm}
\usepackage{color}
\usepackage[mathlines]{lineno}
\usepackage[detect-weight=true, detect-family=true]{siunitx}

\usepackage{hyperref}
\hypersetup{colorlinks=false}

\begin{document}

\preprint{AIP/123-QED}

\newcommand{\RomanNumeralCaps}[1]
    {\MakeUppercase{\romannumeral #1}}

\title{Magnetic field resilient superconducting coplanar waveguide resonators \\for hybrid cQED experiments}

\author{J. G. Kroll}
\affiliation{ 
	QuTech, Delft University of Technology, Delft, 2600 GA, The Netherlands
}
\affiliation{ 
	Kavli Institute for Nanoscience, Delft University of Technology, Delft, 2600 GA, The Netherlands
}

\author{F. Borsoi}%
\affiliation{ 
	QuTech, Delft University of Technology, Delft, 2600 GA, The Netherlands
}
\affiliation{ 
	Kavli Institute for Nanoscience, Delft University of Technology, Delft, 2600 GA, The Netherlands
}

\author{K. L. van der Enden}
\affiliation{ 
	QuTech, Delft University of Technology, Delft, 2600 GA, The Netherlands
}
\affiliation{ 
	Kavli Institute for Nanoscience, Delft University of Technology, Delft, 2600 GA, The Netherlands
}

\author{W. Uilhoorn}
\affiliation{ 
	QuTech, Delft University of Technology, Delft, 2600 GA, The Netherlands
}
\affiliation{ 
	Kavli Institute for Nanoscience, Delft University of Technology, Delft, 2600 GA, The Netherlands
}

\author{D. de Jong}
\affiliation{ 
	QuTech, Delft University of Technology, Delft, 2600 GA, The Netherlands
}
\affiliation{ 
	Kavli Institute for Nanoscience, Delft University of Technology, Delft, 2600 GA, The Netherlands
}

\author{M. Quintero-P\'erez}
\affiliation{ 
	QuTech, Delft University of Technology, Delft, 2600 GA, The Netherlands
}
\affiliation{Netherlands Organisation for Applied Scientific Research (TNO), Delft, 2600 AD, The Netherlands}

\author{D. J. van Woerkom}
\affiliation{ 
	QuTech, Delft University of Technology, Delft, 2600 GA, The Netherlands
}
\affiliation{ 
	Kavli Institute for Nanoscience, Delft University of Technology, Delft, 2600 GA, The Netherlands
}

\author{A. Bruno}
\affiliation{ 
	QuTech, Delft University of Technology, Delft, 2600 GA, The Netherlands
}
\affiliation{ 
	Kavli Institute for Nanoscience, Delft University of Technology, Delft, 2600 GA, The Netherlands
}

\author{S. R. Plissard}
\thanks{Current address: CNRS, LAAS-CNRS, Université de Toulouse, 31400 Toulouse, France}
\affiliation{ 
Department of Applied Physics, Eindhoven University of Technology, Eindhoven, 5600 MB, The Netherlands
}

\author{D. Car}
\affiliation{ 
Department of Applied Physics, Eindhoven University of Technology, Eindhoven, 5600 MB, The Netherlands
}

\author{E. P. A. M. Bakkers}
\affiliation{ 
Department of Applied Physics, Eindhoven University of Technology, Eindhoven, 5600 MB, The Netherlands
}

\author{M. C. Cassidy}
\thanks{Current address: Microsoft Station Q Sydney, Sydney, NSW 2006, Australia}
\affiliation{ 
	QuTech, Delft University of Technology, Delft, 2600 GA, The Netherlands
}
\affiliation{ 
	Kavli Institute for Nanoscience, Delft University of Technology, Delft, 2600 GA, The Netherlands
}

\author{L. P. Kouwenhoven}
 \email{Leo.Kouwenhoven@Microsoft.com}
\affiliation{ 
	QuTech, Delft University of Technology, Delft, 2600 GA, The Netherlands
}
\affiliation{ 
	Kavli Institute for Nanoscience, Delft University of Technology, Delft, 2600 GA, The Netherlands
}
\affiliation{ 
	Microsoft Station Q Delft, Delft, 2600 GA, The Netherlands
}
\date{\today}

\begin{abstract}
Superconducting coplanar waveguide resonators that can operate in strong magnetic fields are important tools for a variety of high frequency superconducting devices. Magnetic fields degrade resonator performance by creating Abrikosov vortices that cause resistive losses and frequency fluctuations, or suppressing superconductivity entirely. To mitigate these effects we investigate lithographically defined artificial defects in resonators fabricated from NbTiN superconducting films. We show that by controlling the vortex dynamics the quality factor of resonators in perpendicular magnetic fields can be greatly enhanced. Coupled with the restriction of the device geometry to enhance the superconductors critical field, we demonstrate stable resonances that retain quality factors $\simeq 10^5$ at the single photon power level in perpendicular magnetic fields up to $B_\perp \simeq \SI{20}{\milli \tesla}$ and parallel magnetic fields up to $B_\parallel \simeq \SI{6}{\tesla}$. We demonstrate the effectiveness of this technique for hybrid systems by integrating an InSb nanowire into a field resilient superconducting resonator, and use it to perform fast charge readout of a gate defined double quantum dot at $B_\parallel = \SI{1}{\tesla}$.
\end{abstract}

\pacs{Valid PACS appear here}
\maketitle

\section{Introduction}
Superconducting (SC) coplanar waveguide (CPW) resonators are invaluable tools for parametric amplifiers \cite{Castellanos-Beltran2007a, Tholen2007}, photon detectors that operate from the infra-red to X-ray frequencies \cite{Mazin2002, Day2003a, Vardulakis2008} and hybrid systems that couple superconducting circuits to cold atoms \cite{Hattermann2017}, solid state spins ensembles \cite{Kubo2010, Amsuss2011, Ranjan2013}, nanomechanical resonators \cite{Regal2008,Teufel2008} and semiconducting devices \cite{Burkard2006,Petersson2012,Viennot2015,Liu2015,DeLange2015,Larsen2015,Kroll2018}. They are also a key component in a variety of quantum computing (QC) platforms, where they are used for readout, control and long-range interconnection of superconducting \cite{Blais2004, Wallraff2004,Wallraff2005,Majer2007}, semiconducting \cite{Stockklauser2017,Landig2018,Mi2018,Samkharadze2018} and topological \cite{Hyart2013,Plugge2017} qubits. 
\label{motivation}

A common requirement among these schemes is that the SC resonators be low loss in order to minimize unwanted relaxation via the Purcell effect \cite{Koch2007} and maximize two qubit gates fidelities \cite{Harvey}. Significant progress has been made in identifying the major loss mechanisms involved. Abundant defects in amorphous materials that couple dissipatively to the electric field of SC CPW resonators and act as effective two level systems (TLSs) were the first to be identified \cite{Gao2008}. Their detrimental effects can be mitigated by minimizing the interface area and improving the quality of interfaces. Subsequently, stray infrared radiation was shown to generate quasiparticles in the SC, significantly increasing losses and necessitating multi-stage radiative shielding to isolate the resonators from their environment \cite{Barends2011}. In addition, when fabricated from type II superconductors, even small magnetic fields ($B_{\text{c}_1} \simeq \SI{1}{\micro \tesla}$) result in the creation of Abrikosov vortices: regions of supercurrent that circulate a non-superconducting core. When exposed to high frequency radiation the Lorentz force causes them to oscillate, generating quasiparticles and increasing losses \cite{Song2011}. Encasing the resonators in multiple layers of magnetic shielding strongly suppresses the number of vortices generated by local magnetic fields, while the use of a type I SC such as Al allows for complete expulsion of magnetic flux via the Meissner effect. Through a combination of these techniques, internal quality factors ($Q_\text{i}$) in excess of $10^6$ have been demonstrated at the single photon power levels \cite{Megrant2012,Bruno2015}.

Many semiconducting and topological QC schemes are accompanied by an additional requirement: strong magnetic fields that approach and sometimes exceed $B = \SI{1}{\tesla}$. This renders traditional magnetic shielding methods useless and destroys the superconductivity of the Al commonly used to fabricate SC CPW resonators ($B_c \simeq $ \SI{10}{\milli \tesla} for bulk Al). One possible solution utilizes type II superconductors with high upper critical magnetic fields such as MoRe, TiN or NbTiN, allowing the superconductivity to persist in very strong magnetic fields ($B^\text{MoRe}_{\text{c}_2} > \SI{8}{\tesla}$ \cite{Calado2015}, $B^\text{NbTiN}_{\text{c}_2} > \SI{9}{\tesla}$ \cite{VanWoerkom2015}) while possessing a low enough density of TLSs to allow high $Q_\text{i}$ resonators at zero field \cite{Vissers2010,Singh2014}. In order to create high $Q_\text{i}$ SC resonators that can survive in strong magnetic fields, controlling the creation and dynamics of Abrikosov vortices is key. Previous studies have utilized the intrinsic disorder in NbTiN \cite{Kwon2018} and YBCO \cite{YBCO} films, or lithographically defined artificial defect sites \cite{Bothner2011} to pin the vortices, preventing dissipation and demonstrating moderate quality factors up to $10^4$ in parallel magnetic fields. In other studies, the resonator geometry has been restricted below the superconducting penetration depth to prevent the formation of vortices entirely \cite{martinis,kuit,degraaf,Samkharadze2016}. Impressively, nanowire resonators have achieved $Q_\text{i}$ of $10^5$ at $B_{||} \simeq \SI{6}{\tesla}$ and $Q_\text{i} = 10^4$ at $B_\perp = $ \SI{350}{\milli \tesla}, however this non-standard geometry may hinder their implementation into complex systems due to the challenging lithography, the required removal of the ground plane and the strong sensitivity of the resonator frequency to variations in the film's kinetic inductance. When considering reliability and scalability, devices based on CPW resonators that have been rendered field compatible may be preferable.

In this work we demonstrate that thin film NbTiN SC CPW resonators with lithographically defined artificial defect sites (hereafter referred to as `holes') can retain a high $Q_\text{i}$ in strong magnetic fields. We first study how the SC film thickness affects the response and $Q_\text{i}$ of resonators in parallel magnetic fields in order to optimize field resilience. We then determine how the hole density affects the dynamics of Abrikosov vortices, enabling $Q_\text{i} \simeq 10^5$ to be retained in perpendicular magnetic fields up to $B_\perp \simeq \SI{20}{\milli \tesla}$. Combining these results, we reduce the number of vortices that are generated by restricting the film thickness and utilize holes to control the vortices that do occur. This allows $Q_\text{i} \simeq 10^5$ at single photon power levels to be retained up to $B_\parallel \simeq$ \SI{6}{\tesla}. Finally, we use these patterned resonators to perform fast charge readout of a hybrid InSb nanowire double quantum dot device at $B_\parallel = \SI{1}{\tesla}$.

\section{Methods}

\begin{figure}
\includegraphics{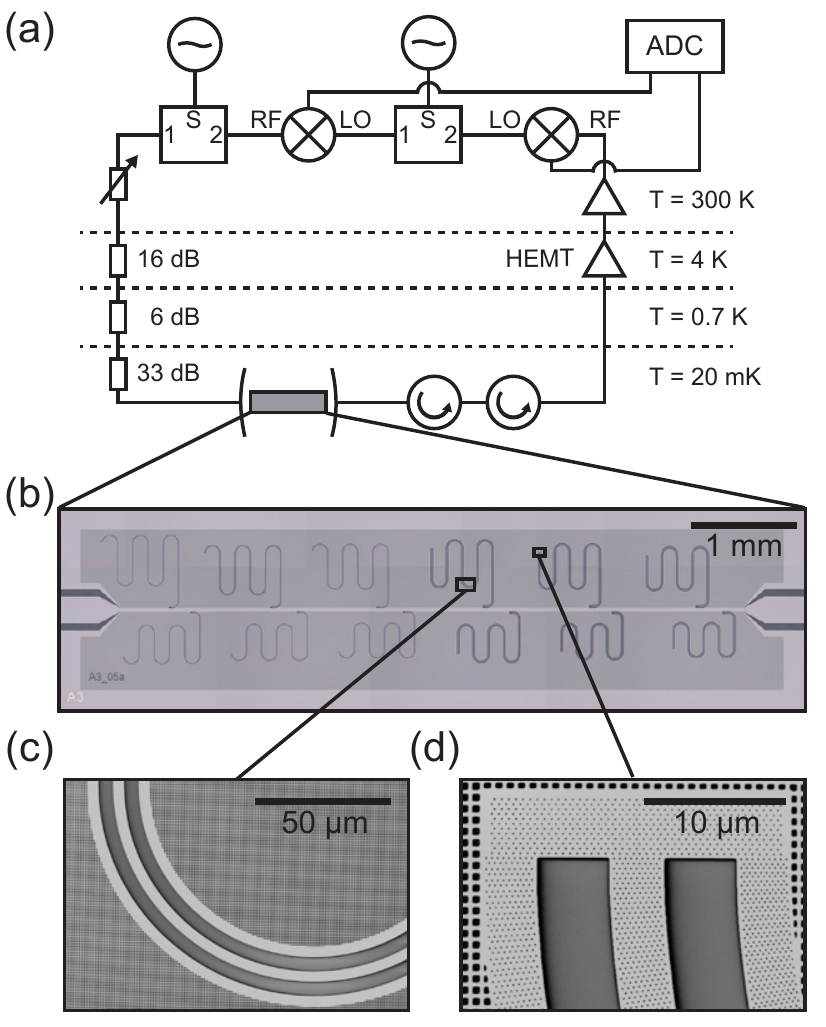}
\caption{(a) Circuit diagram of the heterodyne detection circuit, allowing the complex transmission $S_{21}$ to be measured. (b) A cropped optical image showing a typical device with multiple $\lambda/4$ resonators frequency multiplexed to a common feedline and surrounded by a patterned ground plane. SEM micrographs of a SC CPW resonator without (c) and with holes (d).}
\label{introfig}
\end{figure}

The superconducting CPW resonators are fabricated from NbTiN films sputtered from a 99.99\% purity NbTi target in an Ar/N atmosphere onto 2 inch \SI{430}{\micro \meter} thick sapphire (0001) orientation wafers. A typical \SI{100}{\nano \meter} film has a T$_\text{c}$ of \SI{14}{\kelvin}, a resistivity of \SI{123}{\micro \ohm \centi \meter} and stress of -\SI{400}{\mega \pascal}. The resonators are defined using electron beam lithography (EBL) and a subsequent reactive ion etch (RIE) in an SF$_6$/O$_2$ atmosphere and if desired, holes can be patterned and etched in the same step. After fabrication, the samples are mounted in a light tight copper box and thermally anchored to a dilution refrigerator with a base temperature of \SI{15}{\milli \kelvin}. The complex microwave transmission $S_{21}$ through the sample is measured using standard heterodyne demodulation techniques [Fig.~\ref{introfig}~(a)] which allow the complex transmission $S_{21}$ to be measured as the probe frequency $f$ is varied. The input line to the sample is heavily attenuated, which suppresses thermal noise and allows the cavity photon occupancy to reach below a single photon. An external magnetic field is applied to the sample with a 3-axis 6-1-\SI{1}{\tesla} vector magnet. The circulators used in this study are magnetically shielded to prevent stray fields from interfering with their operation. Two different types of samples are used in this study. The first, used for characterization experiments, contains several $\lambda/4$ resonators multiplexed to a single feedline [Fig.~\ref{introfig}~(b-d)]. This allows for the coupling quality factor $Q_\text{c}$ and internal quality factor $Q_\text{i}$ to be accurately determined by fitting the transmission spectra. In the second geometry, we use a $\lambda /2 $ resonator geometry that is integrated with a nanowire double quantum dot. In this case, we only fit the loaded quality factor $Q_\text{l}$ . All internal quality factors reported in this manuscript correspond to single photon power levels. See Supplementary Information for detailed notes on fitting procedures and device parameters.

\section{Film Thickness Dependence}

We first utilize the critical field enhancement observed in thin strips of type II superconductors \cite{martinis}. Application of a magnetic field along an axis where the thickness $t$ is significantly smaller than the penetration depth $\lambda$ results in complete expulsion of vortices below an enhanced critical field. This enhanced field takes the form: $B_{\text{c}_{1}} = 1.65\Phi_0 /t^2$, where $\Phi_0$ is the magnetic flux quantum and $t$ is the film thickness \cite{martinis,kuit}. Calculating $B_{\text{c}_{1}}$ for a \SI{22}{\nano \meter} film suggests that the first vortex nucleates at $B_{\text{c}_{1}} = \SI{5.17}{\tesla}$ for a perfectly aligned field, an improvement of many orders of magnitude compared to a bulk film.

\begin{figure}
	\includegraphics{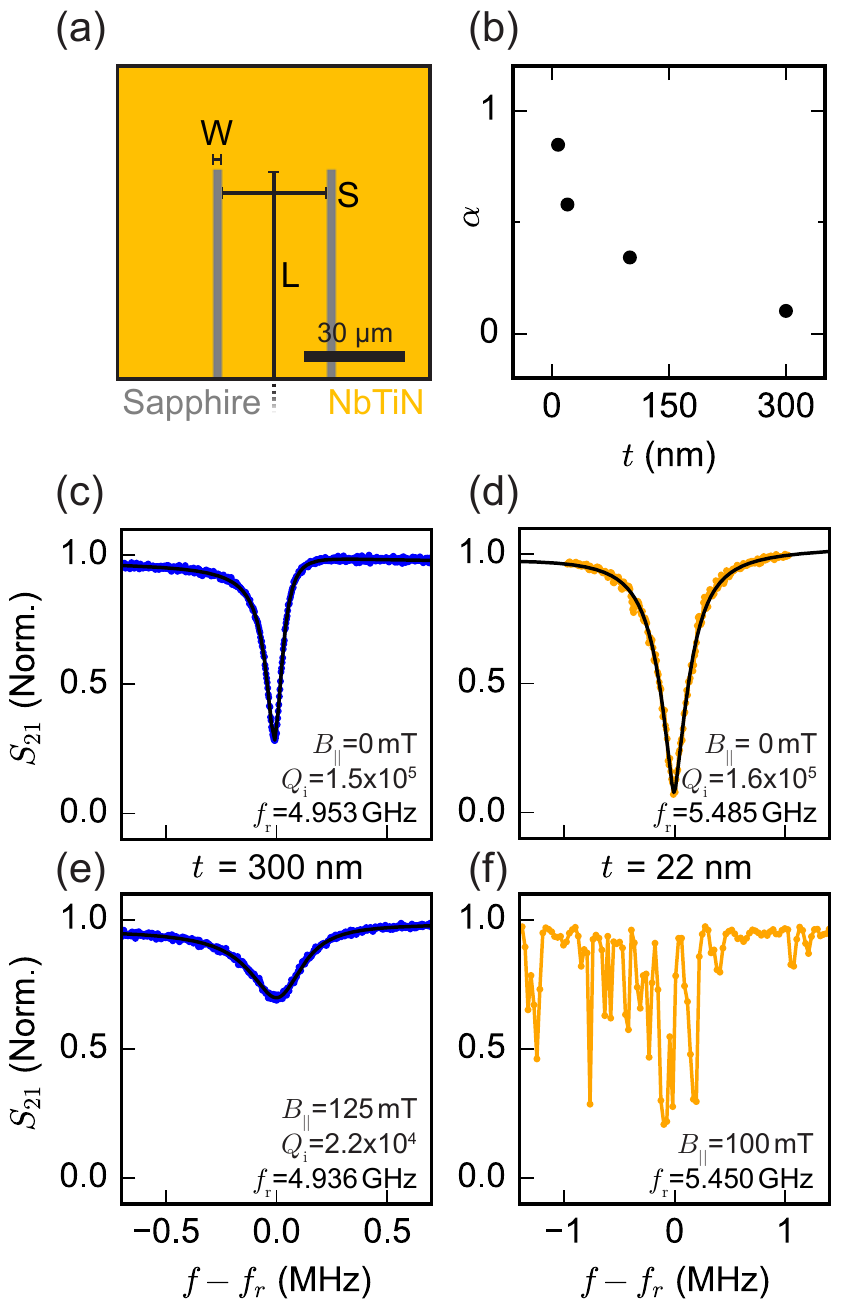}
	\caption{(a) Diagram showing the shorted end of a $\lambda/4$ resonator of length $L$ with a central conductor of width $S$ and gap $W$ and no holes. (b) Extracted value of $\alpha$ for each $t$, showing a sharp increase in $\alpha$ for $t \ll \lambda$. (c,d) $S_{21}$ versus $f$ for resonators with $t = \SI{300}{\nano \meter}$ and $\SI{22}{\nano \meter}$, showing high $Q_\text{i}$, stable resonances at $B = \SI{0}{\milli \tesla}$. (e,f) $S_{21}$ versus $f$ for $t = \SI{300}{\nano \meter}$ and \SI{22}{\nano \meter} at $B_\parallel \ge \SI{100}{\milli \tesla}$.}
	\label{filmthickness}
\end{figure}

To investigate this effect a series of SC CPW resonators without holes were fabricated on 4 different NbTiN films of thicknesses \SI{8}{\nano \meter}, \SI{22}{\nano \meter}, \SI{100}{\nano \meter} and \SI{300}{\nano \meter} [Fig.~\ref{filmthickness}~(a)]. $\lambda$ can be estimated in extremely dirty type II superconductors for $T \rightarrow \SI{0}{\kelvin}$ as $\lambda= \sqrt{\hbar \rho / \pi \mu_0 \Delta_0}$ \cite{Thoen2017} with $\hbar$ being the reduced Planck constant, $\rho$ the film resistivity, $\mu_0$ the permeability of free space and $\Delta_0$ the superconducting gap at $T = \SI{0}{\kelvin}$. For our films $\lambda$ is estimated to vary between 350 to \SI{480}{\nano \meter} depending on the film thickness and measured material parameters.

In thin films, the reduction in Cooper pair density causes the inertial mass of the charge carriers to become important at microwave frequencies, which results in an additional series inductance, or kinetic inductance $L_\text{k}$. We estimate the kinetic inductance fraction $\alpha = L_\text{k}/(L_\text{g}+L_\text{k})$ for each film by comparing the measured resonator frequency to one predicted by an analytical model, where $L_\text{g}$ is the inductance due to the resonator geometry. Thinner films showed an increased kinetic inductance fraction as expected for a strongly disordered superconductor such as NbTiN [Fig.~\ref{filmthickness}~(b)]. While thinner films should increase the magnetic field compatibility, as $\alpha$ approaches 1 the increase in inductance adds additional challenges for impedance matching and frequency targeting. At the same time, frequency targeting of the resonators becomes increasingly difficult due to the sensitivity of $L_\text{k}$ with respect to material parameters that become less consistent in thinner films \cite{Thoen2017}.

At B = \SI{0}{\tesla}, the transmission spectra of each sample shows a series of resonances, each corresponding to a frequency multiplexed resonator. All films display stable resonances with $Q_\text{i}$ varying between $10^4$ and $10^5$. Example resonances for $t$ = \SI{300}{\nano \meter} and \SI{22}{\nano \meter} films are shown in Fig.~\ref{filmthickness}~(c) and (d) respectively.

A magnetic field $B_\parallel$ is then applied parallel to the film using the vector magnet, using the response of the resonator itself to align the magnetic field (see Supplementary Information for alignment details). This results in the different films displaying markedly different behavior. Measurements at $B_{||} > $ \SI{100}{\milli \tesla} on devices with film thicknesses of $t =$ \SI{100}{\nano \meter} and \SI{300}{\nano \meter} [for $t =$ \SI{300}{\nano \meter} see Fig.~\ref{filmthickness}~(e)] show a homogeneous broadening of their resonances, a reduction in their resonance frequency and reduction in $Q_\text{i}$. This is consistent with increased intrinsic losses and increased inductance due to a high density of Abrikosov vortices. At fields above \SI{125}{\milli \tesla}, the reduction in $Q_\text{i}$ and broadening are so extreme that we are unable to resolve the resonances. \label{fieldalignmentprocess}

In contrast, resonators with film thickness of $t$ = \SI{8}{\nano \meter} and \SI{22}{\nano \meter} show inhomogenous broadening [for $t =$ \SI{22}{\nano \meter} see Fig.~\ref{filmthickness}~(f)], with fluctuations of a high $Q$ resonance on the timescale of seconds. We attribute this behavior to a smaller number of vortices being pinned in the superconductor, with vortex depinning events occurring on the same timescale as the measurement. These depinning events may also occur more often in thinner films due to a reduced superconducting order parameter $\Delta_{\text{SC}}$ (see Supplementary Information) and increased variability in $t$. While the reduced vortex number is consistent with an enhancement of the critical field, we note that it is still far below the expected theoretical field enhancement of $B_{\text{c}_1}$ for these films. We attribute this to small misalignments and local deviations in the magnetic field that suppress the critical field enhancement, resulting in a higher than expected vortex density. If resonator performance is to be protected in strong magnetic fields, the detrimental effects of these vortices must be mitigated.

\section{Holes}

\subsection{Perpendicular Field Dependence}
Defects in the film locally reduce $\Delta_{\text{SC}}$ in the superconductor, creating a potential well in which the energy cost required to break the superconductivity in that region is reduced, providing an additional mechanism for controlling Abrikosov vortices. As it is energetically favourable for the vortex to sit in the defect, it experiences a corresponding restoring force pinning it into the defect, minimising losses \cite{Song2011,Bothner2011,plourde,ratchet}. The defects used in this experiment are holes in the superconducting film that are defined by EBL and RIE. This allows them to be patterned in the same step as the resonators and have a diameter smaller than the vortices ($\lambda > d > \xi$), allowing high hole densities $\rho_{\text{h}}$ to be achieved.

We fabricated a sample from a $t = \SI{22}{\nano \meter}$ film with resonators of different $\rho_{\text{h}}$ varying from 0 to \SI{28.8}{\per \micro \meter \squared} to determine the effect of holes on resonator performance. The resonators retain a constant CPW geometry and the ground plane is patterned with large square holes that trap residual local magnetic fluxes in the environment (see Supplementary Information). The holes have diameters $d = \SI{100}{\nano \meter}$ and are patterned into hexagonal arrays on the central conductor and edges of the adjacent ground planes. Unlike previous experiments with holes at the edge of the resonator \cite{Bothner2011}, we leave a gap of at least \SI{1}{\micro \meter} to the film edge to avoid interfering with the bulk of the current that flows at the edges of the CPW \cite{Bothner2012}. The hole densities can be converted to a threshold `critical field' $B_{\text{Th}}$ (ranging between 0 and \SI{59.69}{\milli \tesla}) that when applied perpendicular to the plane of the film fills each hole with a single Abrikosov vortex. Above $B_{\text{Th}}$ additional vortices are no longer strongly pinned by the holes but instead only weakly pinned by film defects and interstitial pinning effects.

\begin{figure}
	\includegraphics{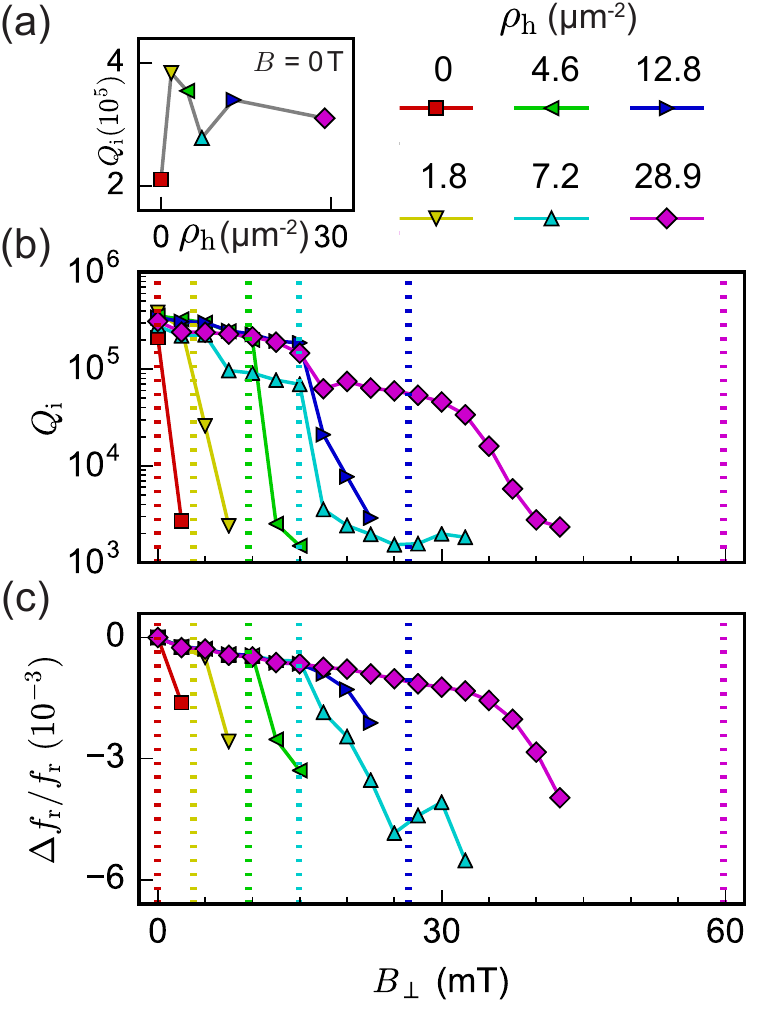}
	\caption{(a) $Q_\text{i}$ versus $\rho_{\text{h}}$ for field-cooled resonators at $B_\perp = \SI{0}{\tesla}$. (b) $Q_\text{i}$ versus $B_{\perp}$ for varying $\rho_\text{h}$. $B_{\text{Th}}$ for each resonator is plotted with a colour matched vertical line. (c) $\Delta f_\text{r}/f_\text{r}$ versus $B_{\perp}$ for varying $\rho_\text{h}$ with $B_{\text{Th}}$ plotted as in (b).}
	\label{perpendicularfieldresponse}

\end{figure}

Field-cooled measurements are performed by applying a perpendicular magnetic field $B_{\perp}$ to the sample above $T_\text{c} \sim \SI{14}{\kelvin}$ of the NbTiN, then cooling the sample to base temperature before performing transmission measurements to characterize the resonators. At $B = \SI{0}{\tesla}$, all resonators are measured with $Q_\text{i} > 10^5$ [Fig.~\ref{perpendicularfieldresponse}~(a)]. Resonators with a non-zero $\rho_{\text{h}}$ trap the few vortices that form due to the local magnetic environment near the CPW, enhancing $Q_\text{i}$. At higher $\rho_{\text{h}}$, $Q_\text{i}$ reduces, possibly due to increased losses occurring in the larger metal-vacuum interface area \cite{Woods2018}.
	
The behavior of $Q_\text{i}$ versus $B_{\perp}$ field is plotted in Fig.~\ref{perpendicularfieldresponse}~(b) together with the calculated $B_{\text{Th}}$ for each hole density. For $B < B_\text{Th}$, each resonator retains a high $Q_\text{i}$ as all vortices can be trapped in holes. Once every hole is full ($B > B_{\text{Th}}$) $Q_\text{i}$ sharply decreases, as additional vortices are then only weakly pinned and can become itinerant in the SC film, resulting in resistive losses. The same response is observed in the fractional frequency $\Delta f_\text{r}/f_\text{r} = (f_\text{r}-f^{B=0}_\text{r})/f_\text{r}$ of the resonators [Fig.~\ref{perpendicularfieldresponse}~(c)]. For $B < B_{\text{Th}}$, the magnetic field increases the rate of Cooper pair breaking, increasing $L_\text{k}$ and resulting in a gradual reduction of $f_\text{r}$ \cite{Samkharadze2016}. Once $B > B_{\text{Th}}$ the vortices are no longer pinned by the hole lattice and become itinerant, significantly increasing the inductive load on the resonator resulting in a comparatively large frequency shift.

At larger $\rho_{\text{h}}$ ($\geq$ \SI{12.8}{\per \micro\meter \squared}) the results start to deviate from the expected response. We attribute this to the vortex density approaching a regime where the vortices start to overlap, meaning that the superconductivity can no longer assumed to be homogeneous, increasing losses. Unusually, the resonator with highest $\rho_\text{h}$ (i.e. the `hole-iest' resonator) experiences a reduction in $Q_\text{i}$ at $\sim \SI{15}{\milli \tesla}$ along with the other resonators, but retains a relatively high $Q_\text{i} \simeq 10^5$ and flat $\Delta f_\text{r}/f_\text{r}$ response until $B_\perp \simeq \SI{35}{\milli \tesla}$. This behavior is consistent with a fraction of the vortices penetrating the SC film, but not being itinerant. This would result in the observed $Q_\text{i}$ reduction while protecting the resonators $f_\text{r}$ from the sharp reduction observed in the other resonators. This is suggestive of interstitial pinning effects that have been extensively studied previously and could be used to further increase the perpendicular field compatibility of these resonators \cite{interstitial1,interstitial2,interstitial3}.

\subsection{Parallel Field Dependence}

\begin{figure}
	\includegraphics{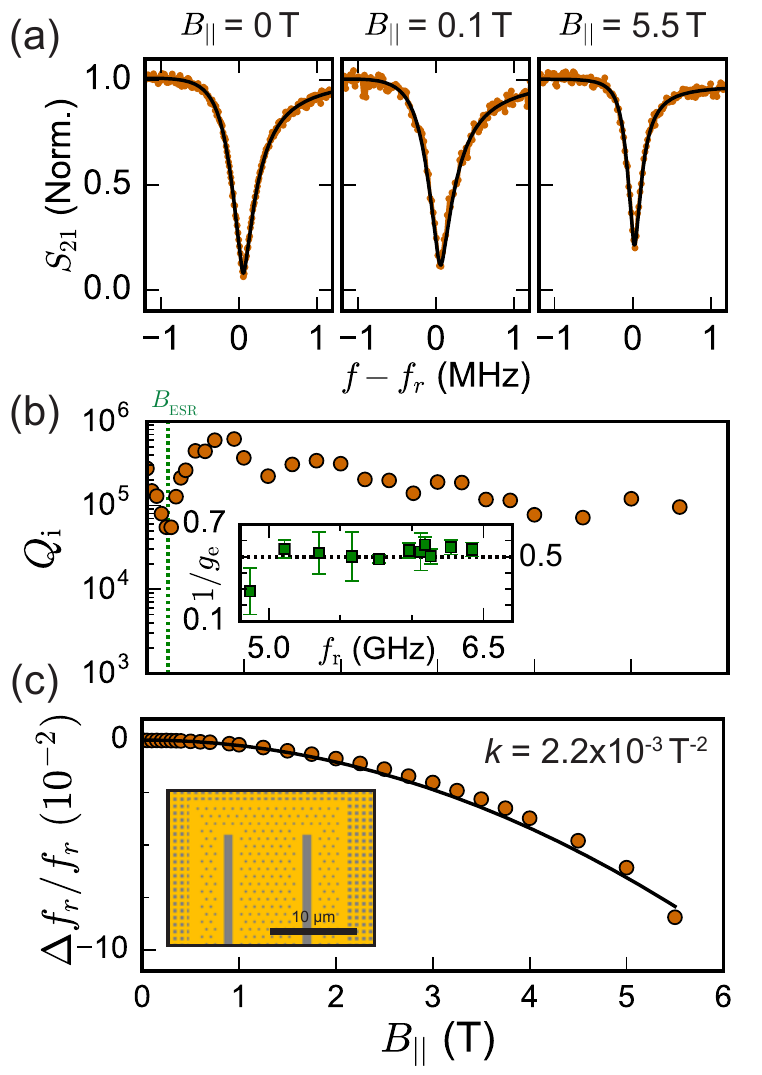}
	\caption{(a) $S_{21}$ versus $f-f_\text{r}$ for a typical $t = \SI{22}{\nano \meter}$ resonator with holes at $B_\parallel =$ \SI{0}{\tesla}, \SI{0.1}{\tesla} and \SI{5.5}{\tesla}. (b) Extracted $Q_\text{i}$ versus $B_\parallel$. Inset shows the inverse of the extracted $g_\text{e}$-factor from $Q_\text{i}$ minima of each resonator as a function of $B_{||}$. (c) Extracted $\Delta f_\text{r}/f_\text{r}$ versus $B_\parallel$. A fit of the data with $\Delta f_\text{r}/f_\text{r} = -k B_{||}^2$ gives $k$ = \SI{2.2e-3}{\per \tesla \squared}. Inset is a diagram showing the distribution of holes around the CPW.}
	\label{parallelfieldresponse}
\end{figure}

To demonstrate that thin films and a low density of holes can be used to increase the resilience of resonators to strong parallel magnetic fields, a sample with 12 resonators of varying CPW dimensions but constant hole density were fabricated from $t$ = \SI{22}{\nano \meter} films. The holes are \SI{300}{\nano \meter} in diameter and arranged in hexagonal lattice around the CPW with a density of $\rho_{\text{h}} = \SI{1.2}{\per \micro \meter \squared}$ [Fig.~\ref{parallelfieldresponse}~(c) inset]. The device is then cooled at $B = \SI{0}{\tesla}$ and $S_{21}$ measurements performed to determine the $Q_\text{i}$ of all 12 resonators as a function of $B_{||}$, with $B_{||}$ applied as in Sec.~\ref{fieldalignmentprocess} using the response of the resonator for field alignment. 

$S_{21}$ versus $f-f_\text{r}$ for an example resonance at $B_\parallel =$ \SI{0}{\tesla}, \SI{0.1}{\tesla} and \SI{5.5}{\tesla} [Fig.~\ref{parallelfieldresponse}~(a)] shows a stable, high $Q_\text{i}$ peak at all fields, a key requirement for their effective use in superconducting circuits. Fitting the resonator response to extract $Q_\text{i}$ versus $B_\parallel$ [Fig.~\ref{parallelfieldresponse}~(b)] reveals that $Q_\text{i} \sim 10^5$ can be retained up to $B_\parallel =$ \SI{5.5}{\tesla}. A dip in $Q_\text{i}$ is observed at $B_\text{ESR} \sim$ \SI{200}{\milli \tesla} and is attributed to coupling to an electron spin resonance that increases losses in the cavity. This feature is observed in all resonators, with the frequency dependence corresponding to paramagnetic impurities with a Land\'{e} $g$-factor of $\sim$ 2 [Fig.~\ref{parallelfieldresponse}~(b) inset] as observed in \cite{Samkharadze2016}. 

As $B_\parallel$ is increased, the resonator $f_\text{r}$ reduces [Fig.~\ref{parallelfieldresponse}~(c)] due to the increased rate of Cooper pair breaking from the applied $B_{||}$. It can be modelled as a parabolic decrease in $B_{||}$ as $\Delta f_\text{r}/f_\text{r} = -kB_{||}^2$, with $k = \frac{\pi}{48}[t^2e^2D/\hbar k_\text{B} T_{\text{c}}]$ dependent on $t$, $T_{\text{c}}$ and $D$ the electron diffusion constant in NbTiN. With all other parameters known, we determine $D = \SI{0.79}{\centi \meter \squared \per \second}$, consistent with previous results \cite{Samkharadze2016}. The field resilience up to $B_\parallel \simeq \SI{6}{\tesla}$ is observed in all resonators, implying that this is a generic recipe by which field resilient SC CPW resonators can be fabricated (see the Supplementary Information for measurements of all 12 resonators).

\section{Charge readout of a hybrid I\lowercase{n}S\lowercase{b} nanowire device at \SI{1}{\tesla}}

\begin{figure}
	\includegraphics{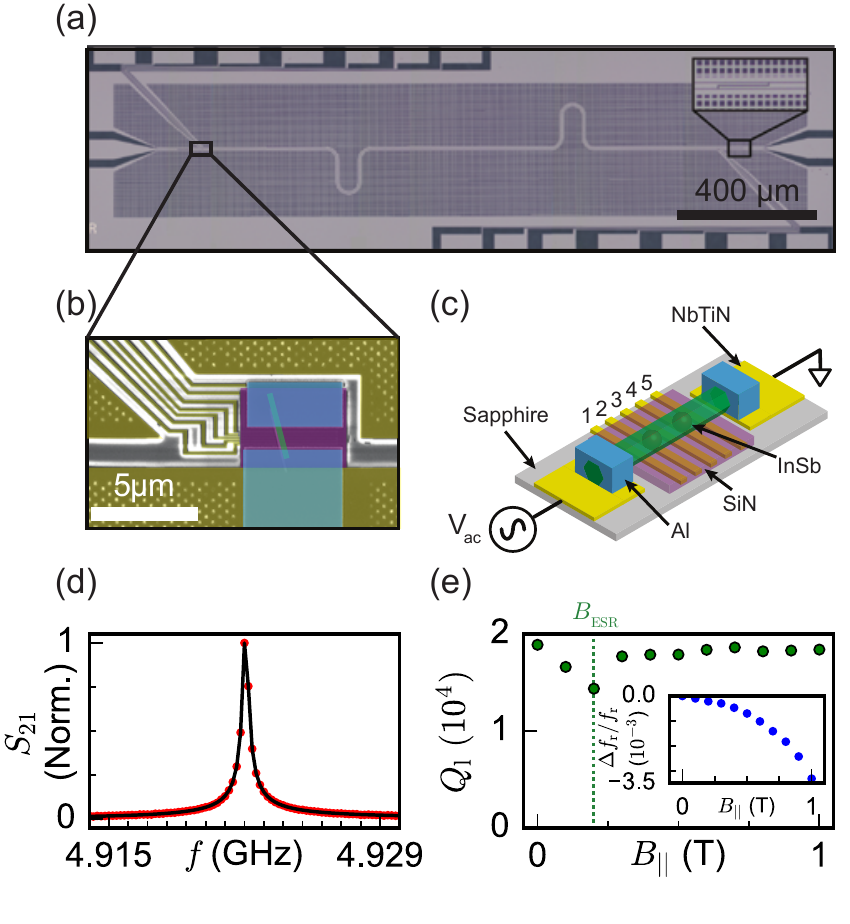}
	\caption{(a) Optical microscope image (b) scanning electron microscope image and (c) 3D diagram of the hybrid device. At each end of the cavity, a single InSb nanowire is deposited on local fine gates. Ti/Al is used to contact the nanowire to the resonator's central conductor and ground. Local fine gates (1-5) are used to electrostatically define the DQD. Coupling capacitors control the photon lifetime in the cavity. (d) $S_{21}$ versus $f$ shows a single resonance at $f = \SI{4.922}{\giga \hertz}$ with $Q_\text{l} = 1.8$ x $10^4$. (e) $Q_{\text{l}}$ as a function of $B_{||}$ demonstrates the resonator is unaffected by strong magnetic fields. Inset is $\Delta f_\text{r}/f_\text{r}$ versus $B_{\parallel}$ displaying a small parabolic frequency shift.}
	\label{hybrid_device}
\end{figure}
Double quantum dots (DQDs) and high quality factor SC CPW resonators in combination form a superconductor-semiconductor `hybrid' system allowing for the high bandwidth sensing of DQD charge configurations in a variety of material systems using techniques from circuit quantum electrodynamics (cQED) \cite{Frey,Viennot2015,Ranjan2015,Wang2016}. More exotic hybrid systems such as spin qubits \cite{Petersson2012,Mi2018}, Majorana box qubits \cite{Plugge2017} and spin-transmon hybrids also require the application of strong parallel magnetic fields, a condition that to date has proven challenging for traditional SC CPW resonators. To demonstrate the importance of our field resilient patterned SC CPW for hybrid systems we integrate a pair of InSb nanowires into a $\lambda/2$ resonator and perform fast charge readout of DQDs at a magnetic field of $B_\parallel = \SI{1}{\tesla}$.

The $\lambda/2$ resonator, holes and the electrostatic gates required for forming the DQDs are formed in a single lithography step followed by SF$_6$/O$_2$ reactive ion etch of a \SI{20}{\nano \meter} NbTiN film [see Fig.~\ref{hybrid_device}~(a)]. The holes ($d = \SI{150}{\nano \meter}$) are arranged in a hexagonal lattice with a density of $\rho_{\text{h}} = \SI{3.2}{\per \micro \meter \squared}$. At each end of the resonator at the electric field maxima, we deposit a single nanowire on top of a set of 5 fine gates [Fig.~\ref{hybrid_device}~(a-c)]. A \SI{30}{\nano \meter} layer of sputtered SiN$_x$ electrically isolates the nanowire from the underlying gates. Following sulfur etching to remove the surface oxide from the InSb nanowire \cite{Gul2017}, \SI{150}{\nano \meter} Ti/Al contacts are evaporated on each end of the nanowire to define the electrical potential in the nanowire. One contact is connected directly to the central conductor of the resonator, and the other end directly to the ground plane to enhance the coupling between the DQD and the cavity's electric field \cite{Petersson2012}. By applying voltages to the electrostatic gates, we can control the electron occupation in the nanowire, either turning the device off, or tuning it into a single or double quantum dot configuration.

Transmission measurements of the resonator with each nanowire depleted show a single resonance at frequency $f_\text{r} = \SI{4.922}{\giga \hertz}$ with loaded quality factor $Q_\text{l} \simeq 1.8$x$10^5$ [Fig.~\ref{hybrid_device}~(d)]. The resonator is in the over-coupled regime, with the coupling capacitances of the input and output ports controlling the photon lifetime in the cavity. 

The magnetic field $B_{||}$ is applied parallel to the plane of the resonator [Fig.~\ref{hybrid_device}~(e)], using the response of the resonator itself to align the field (as described in Sec.~\ref{fieldalignmentprocess}). Similarly to the $\lambda/4$ resonators, $f_\text{r}$ reduces parabolically in $B_{||}$ due to the increased $L_{\text{k}}$ and $Q_\text{l}$ experiences a decrease at $B_\text{ESR} \simeq \SI{200}{\milli \tesla}$, corresponding to paramagnetic impurities coupling to the cavity via electron spin resonance [Fig.~\ref{hybrid_device}~(e) and inset]. 

\begin{figure}
	\includegraphics{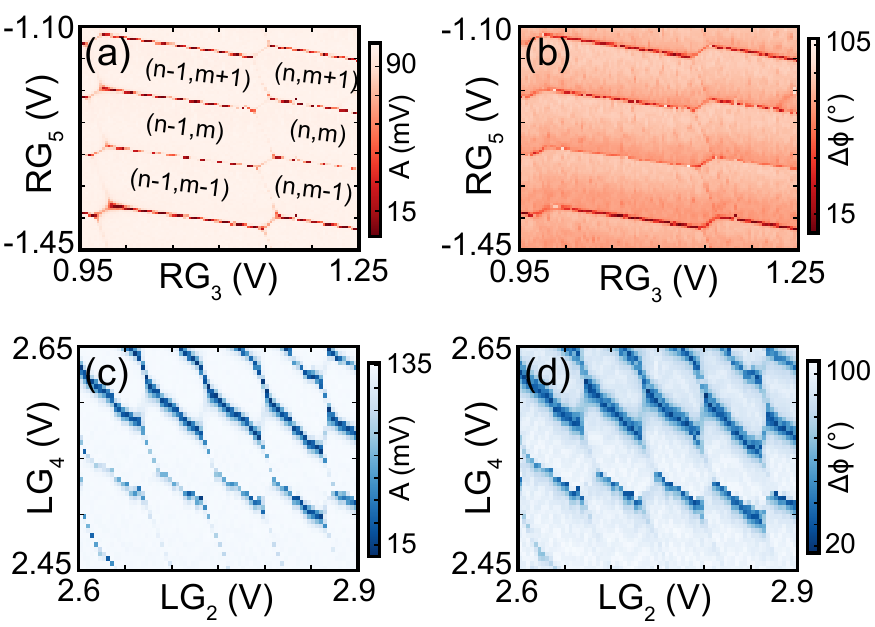}
	\caption{Measuring $S_{21}$ at $f_\text{r}$ at $B = \SI{1}{\tesla}$ as $V$ on the fine gates is varied generates charge stability diagrams of the right (a,b) and left (c,d) nanowire DQDs. The electron occupation $n$ ($m$) of the left (right) QD can be modified by tuning the gate voltages.}
	\label{hybrid_device2}
\end{figure}

Once the external magnetic field is aligned and has reached $B_{||} = \SI{1}{\tesla}$, the resonator can be used for tuning the charge occupancy in each of the nanowire DQDs without DC transport. The DQDs are formed by monitoring the microwave response of the resonator and using an applied voltage on outer gates 1 and 5 [Fig.~\ref{hybrid_device}~(c)] to define a single QD, then using the middle gate 3 to form a DQD. Electron transitions between the dot and lead are a dissipative process that cause a reduction in $Q_{\text{l}}$, which can be observed as a reduction in the $S_{21}$ peak amplitude. Additionally, as the inter-dot tunnel coupling rate of the electrons $t_\text{c}$ approaches $f_\text{r}$, the transition coherently couples to the electric field in the cavity via a Jaynes-Cummings type interaction. This results in a repulsion of the resonator, shifting the resonator frequency and changing the measured phase $\phi$ \cite{Frey}.

Monitoring the amplitude and phase fluctuations as the plunger gates are varied allows the dot-lead and inter-dot transitions to be identified and the charge stability diagrams for each of the DQDs to be measured at $B_{||} = \SI{1}{\tesla}$ [Fig.~\ref{hybrid_device2}]. For the right DQD, gates 2, 4 and 5 are used to define the dots while 3 and 5 control the electron occupancy [Fig.~\ref{hybrid_device2}~(a,b)]. In the left DQD the electrical confinement is defined by 1, 3, 5, and the charge occupancy controlled by 2 and 4 [Fig.~\ref{hybrid_device2}~(c,d)]. For each DQD, the charge occupancy in the left (right) dot is denoted with $n$ ($m$).

The measurements were performed at $B_{||} = \SI{1}{\tesla}$, an order of magnitude improvement over previous studies \cite{Petersson2012,Samkharadze2018}, and a magnetic field of relevance for numerous topological quantum computing proposals \cite{Hyart2013,Plugge2017} and the investigation of mesoscopic phenomena in hybrid systems under strong magnetic fields \cite{Kroll2018}.

\section{Conclusion}

These results indicate that by controlling vortex dynamics in SC CPW resonators a $Q_\text{i} \simeq 10^5$ can be retained in a perpendicular magnetic field of $B_\perp \simeq \SI{20}{\milli \tesla}$. When combined with film thickness reduction to enhance $B_{\text{c}_1}$, we establish a reliable fabrication recipe to create SC CPW resonators that can retain a high $Q_\text{i} \sim 10^5$ in strong parallel magnetic fields up to $B_\parallel \simeq \SI{6}{\tesla}$. We demonstrate the importance of these techniques for hybrid systems by coupling a superconducting resonator to DQDs electrostatically defined in InSb nanowires. Using high frequency measurement techniques we demonstrate device operation and determine the charge stability diagrams of two DQDs at $B_\parallel = \SI{1}{\tesla}$, a magnetic field of relevance for mesoscopic physics studies and semiconducting, topological and hybrid quantum computing schemes.

\begin{acknowledgments}
We thank N. Samkharadze, C. Dickel, F. L\"uthi, N. Langford and L. DiCarlo for useful discussions. This work has been supported by the Netherlands Organisation for Scientific Research (NWO/OCW), as part of the Frontiers of Nanoscience (NanoFront) program, the European Research Council (ERC) and Microsoft Corporation Station Q.
\end{acknowledgments}

\nocite{*}


\begin{thebibliography}{58}%
	\makeatletter
	\providecommand \@ifxundefined [1]{%
		\@ifx{#1\undefined}
	}%
	\providecommand \@ifnum [1]{%
		\ifnum #1\expandafter \@firstoftwo
		\else \expandafter \@secondoftwo
		\fi
	}%
	\providecommand \@ifx [1]{%
		\ifx #1\expandafter \@firstoftwo
		\else \expandafter \@secondoftwo
		\fi
	}%
	\providecommand \natexlab [1]{#1}%
	\providecommand \enquote  [1]{``#1''}%
	\providecommand \bibnamefont  [1]{#1}%
	\providecommand \bibfnamefont [1]{#1}%
	\providecommand \citenamefont [1]{#1}%
	\providecommand \href@noop [0]{\@secondoftwo}%
	\providecommand \href [0]{\begingroup \@sanitize@url \@href}%
	\providecommand \@href[1]{\@@startlink{#1}\@@href}%
	\providecommand \@@href[1]{\endgroup#1\@@endlink}%
	\providecommand \@sanitize@url [0]{\catcode `\\12\catcode `\$12\catcode
		`\&12\catcode `\#12\catcode `\^12\catcode `\_12\catcode `\%12\relax}%
	\providecommand \@@startlink[1]{}%
	\providecommand \@@endlink[0]{}%
	\providecommand \url  [0]{\begingroup\@sanitize@url \@url }%
	\providecommand \@url [1]{\endgroup\@href {#1}{\urlprefix }}%
	\providecommand \urlprefix  [0]{URL }%
	\providecommand \Eprint [0]{\href }%
	\providecommand \doibase [0]{http://dx.doi.org/}%
	\providecommand \selectlanguage [0]{\@gobble}%
	\providecommand \bibinfo  [0]{\@secondoftwo}%
	\providecommand \bibfield  [0]{\@secondoftwo}%
	\providecommand \translation [1]{[#1]}%
	\providecommand \BibitemOpen [0]{}%
	\providecommand \bibitemStop [0]{}%
	\providecommand \bibitemNoStop [0]{.\EOS\space}%
	\providecommand \EOS [0]{\spacefactor3000\relax}%
	\providecommand \BibitemShut  [1]{\csname bibitem#1\endcsname}%
	\let\auto@bib@innerbib\@empty
	\bibitem [{\citenamefont {Castellanos-Beltran}\ and\ \citenamefont
		{Lehnert}(2007)}]{Castellanos-Beltran2007a}%
	\BibitemOpen
	\bibfield  {author} {\bibinfo {author} {\bibfnamefont {M.~A.}\ \bibnamefont
			{Castellanos-Beltran}}\ and\ \bibinfo {author} {\bibfnamefont {K.~W.}\
			\bibnamefont {Lehnert}},\ }\href {\doibase 10.1063/1.2773988} {\bibfield
		{journal} {\bibinfo  {journal} {Applied Physics Letters}\ }\textbf {\bibinfo
			{volume} {91}},\ \bibinfo {pages} {083509} (\bibinfo {year}
		{2007})}\BibitemShut {NoStop}%
	\bibitem [{\citenamefont {Thol{\'{e}}n}\ \emph {et~al.}(2007)\citenamefont
		{Thol{\'{e}}n}, \citenamefont {Erg{\"{u}}l}, \citenamefont {Doherty},
		\citenamefont {Weber}, \citenamefont {Gr{\'{e}}gis},\ and\ \citenamefont
		{Haviland}}]{Tholen2007}%
	\BibitemOpen
	\bibfield  {author} {\bibinfo {author} {\bibfnamefont {E.~A.}\ \bibnamefont
			{Thol{\'{e}}n}}, \bibinfo {author} {\bibfnamefont {A.}~\bibnamefont
			{Erg{\"{u}}l}}, \bibinfo {author} {\bibfnamefont {E.~M.}\ \bibnamefont
			{Doherty}}, \bibinfo {author} {\bibfnamefont {F.~M.}\ \bibnamefont {Weber}},
		\bibinfo {author} {\bibfnamefont {F.}~\bibnamefont {Gr{\'{e}}gis}}, \ and\
		\bibinfo {author} {\bibfnamefont {D.~B.}\ \bibnamefont {Haviland}},\ }\href
	{\doibase 10.1063/1.2750520} {\bibfield  {journal} {\bibinfo  {journal}
			{Applied Physics Letters}\ }\textbf {\bibinfo {volume} {90}},\ \bibinfo
		{pages} {253509} (\bibinfo {year} {2007})}\BibitemShut {NoStop}%
	\bibitem [{\citenamefont {Mazin}\ \emph {et~al.}(2002)\citenamefont {Mazin},
		\citenamefont {Day}, \citenamefont {LeDuc}, \citenamefont {Vayonakis},\ and\
		\citenamefont {Zmuidzinas}}]{Mazin2002}%
	\BibitemOpen
	\bibfield  {author} {\bibinfo {author} {\bibfnamefont {B.~A.}\ \bibnamefont
			{Mazin}}, \bibinfo {author} {\bibfnamefont {P.~K.}\ \bibnamefont {Day}},
		\bibinfo {author} {\bibfnamefont {H.~G.}\ \bibnamefont {LeDuc}}, \bibinfo
		{author} {\bibfnamefont {A.}~\bibnamefont {Vayonakis}}, \ and\ \bibinfo
		{author} {\bibfnamefont {J.}~\bibnamefont {Zmuidzinas}}\ }(\bibinfo
	{publisher} {International Society for Optics and Photonics},\ \bibinfo
	{year} {2002})\ p.\ \bibinfo {pages} {283}\BibitemShut {NoStop}%
	\bibitem [{\citenamefont {Day}\ \emph {et~al.}(2003)\citenamefont {Day},
		\citenamefont {LeDuc}, \citenamefont {Mazin}, \citenamefont {Vayonakis},\
		and\ \citenamefont {Zmuidzinas}}]{Day2003a}%
	\BibitemOpen
	\bibfield  {author} {\bibinfo {author} {\bibfnamefont {P.~K.}\ \bibnamefont
			{Day}}, \bibinfo {author} {\bibfnamefont {H.~G.}\ \bibnamefont {LeDuc}},
		\bibinfo {author} {\bibfnamefont {B.~A.}\ \bibnamefont {Mazin}}, \bibinfo
		{author} {\bibfnamefont {A.}~\bibnamefont {Vayonakis}}, \ and\ \bibinfo
		{author} {\bibfnamefont {J.}~\bibnamefont {Zmuidzinas}},\ }\href {\doibase
		10.1038/nature02037} {\bibfield  {journal} {\bibinfo  {journal} {Nature}\
		}\textbf {\bibinfo {volume} {425}},\ \bibinfo {pages} {817} (\bibinfo {year}
		{2003})}\BibitemShut {NoStop}%
	\bibitem [{\citenamefont {Vardulakis}\ \emph {et~al.}(2008)\citenamefont
		{Vardulakis}, \citenamefont {Withington}, \citenamefont {Goldie},\ and\
		\citenamefont {Glowacka}}]{Vardulakis2008}%
	\BibitemOpen
	\bibfield  {author} {\bibinfo {author} {\bibfnamefont {G.}~\bibnamefont
			{Vardulakis}}, \bibinfo {author} {\bibfnamefont {S.}~\bibnamefont
			{Withington}}, \bibinfo {author} {\bibfnamefont {D.~J.}\ \bibnamefont
			{Goldie}}, \ and\ \bibinfo {author} {\bibfnamefont {D.~M.}\ \bibnamefont
			{Glowacka}},\ }\href {\doibase 10.1088/0957-0233/19/1/015509} {\bibfield
		{journal} {\bibinfo  {journal} {Measurement Science and Technology}\ }\textbf
		{\bibinfo {volume} {19}},\ \bibinfo {pages} {015509} (\bibinfo {year}
		{2008})}\BibitemShut {NoStop}%
	\bibitem [{\citenamefont {Hattermann}\ \emph {et~al.}(2017)\citenamefont
		{Hattermann}, \citenamefont {Bothner}, \citenamefont {Ley}, \citenamefont
		{Ferdinand}, \citenamefont {Wiedmaier}, \citenamefont {S{\'{a}}rk{\'{a}}ny},
		\citenamefont {Kleiner}, \citenamefont {Koelle},\ and\ \citenamefont
		{Fort{\'{a}}gh}}]{Hattermann2017}%
	\BibitemOpen
	\bibfield  {author} {\bibinfo {author} {\bibfnamefont {H.}~\bibnamefont
			{Hattermann}}, \bibinfo {author} {\bibfnamefont {D.}~\bibnamefont {Bothner}},
		\bibinfo {author} {\bibfnamefont {L.~Y.}\ \bibnamefont {Ley}}, \bibinfo
		{author} {\bibfnamefont {B.}~\bibnamefont {Ferdinand}}, \bibinfo {author}
		{\bibfnamefont {D.}~\bibnamefont {Wiedmaier}}, \bibinfo {author}
		{\bibfnamefont {L.}~\bibnamefont {S{\'{a}}rk{\'{a}}ny}}, \bibinfo {author}
		{\bibfnamefont {R.}~\bibnamefont {Kleiner}}, \bibinfo {author} {\bibfnamefont
			{D.}~\bibnamefont {Koelle}}, \ and\ \bibinfo {author} {\bibfnamefont
			{J.}~\bibnamefont {Fort{\'{a}}gh}},\ }\href {\doibase
		10.1038/s41467-017-02439-7} {\bibfield  {journal} {\bibinfo  {journal}
			{Nature Communications}\ }\textbf {\bibinfo {volume} {8}},\ \bibinfo {pages}
		{2254} (\bibinfo {year} {2017})}\BibitemShut {NoStop}%
	\bibitem [{\citenamefont {Kubo}\ \emph {et~al.}(2010)\citenamefont {Kubo},
		\citenamefont {Ong}, \citenamefont {Bertet}, \citenamefont {Vion},
		\citenamefont {Jacques}, \citenamefont {Zheng}, \citenamefont {Dr{\'{e}}au},
		\citenamefont {Roch}, \citenamefont {Auffeves}, \citenamefont {Jelezko},
		\citenamefont {Wrachtrup}, \citenamefont {Barthe}, \citenamefont {Bergonzo},\
		and\ \citenamefont {Esteve}}]{Kubo2010}%
	\BibitemOpen
	\bibfield  {author} {\bibinfo {author} {\bibfnamefont {Y.}~\bibnamefont
			{Kubo}}, \bibinfo {author} {\bibfnamefont {F.~R.}\ \bibnamefont {Ong}},
		\bibinfo {author} {\bibfnamefont {P.}~\bibnamefont {Bertet}}, \bibinfo
		{author} {\bibfnamefont {D.}~\bibnamefont {Vion}}, \bibinfo {author}
		{\bibfnamefont {V.}~\bibnamefont {Jacques}}, \bibinfo {author} {\bibfnamefont
			{D.}~\bibnamefont {Zheng}}, \bibinfo {author} {\bibfnamefont
			{A.}~\bibnamefont {Dr{\'{e}}au}}, \bibinfo {author} {\bibfnamefont {J.~F.}\
			\bibnamefont {Roch}}, \bibinfo {author} {\bibfnamefont {A.}~\bibnamefont
			{Auffeves}}, \bibinfo {author} {\bibfnamefont {F.}~\bibnamefont {Jelezko}},
		\bibinfo {author} {\bibfnamefont {J.}~\bibnamefont {Wrachtrup}}, \bibinfo
		{author} {\bibfnamefont {M.~F.}\ \bibnamefont {Barthe}}, \bibinfo {author}
		{\bibfnamefont {P.}~\bibnamefont {Bergonzo}}, \ and\ \bibinfo {author}
		{\bibfnamefont {D.}~\bibnamefont {Esteve}},\ }\href {\doibase
		10.1103/PhysRevLett.105.140502} {\bibfield  {journal} {\bibinfo  {journal}
			{Physical Review Letters}\ }\textbf {\bibinfo {volume} {105}},\ \bibinfo
		{pages} {140502} (\bibinfo {year} {2010})}\BibitemShut {NoStop}%
	\bibitem [{\citenamefont {Ams{\"{u}}ss}\ \emph {et~al.}(2011)\citenamefont
		{Ams{\"{u}}ss}, \citenamefont {Koller}, \citenamefont {N{\"{o}}bauer},
		\citenamefont {Putz}, \citenamefont {Rotter}, \citenamefont {Sandner},
		\citenamefont {Schneider}, \citenamefont {Schramb{\"{o}}ck}, \citenamefont
		{Steinhauser}, \citenamefont {Ritsch}, \citenamefont {Schmiedmayer},\ and\
		\citenamefont {Majer}}]{Amsuss2011}%
	\BibitemOpen
	\bibfield  {author} {\bibinfo {author} {\bibfnamefont {R.}~\bibnamefont
			{Ams{\"{u}}ss}}, \bibinfo {author} {\bibfnamefont {C.}~\bibnamefont
			{Koller}}, \bibinfo {author} {\bibfnamefont {T.}~\bibnamefont
			{N{\"{o}}bauer}}, \bibinfo {author} {\bibfnamefont {S.}~\bibnamefont {Putz}},
		\bibinfo {author} {\bibfnamefont {S.}~\bibnamefont {Rotter}}, \bibinfo
		{author} {\bibfnamefont {K.}~\bibnamefont {Sandner}}, \bibinfo {author}
		{\bibfnamefont {S.}~\bibnamefont {Schneider}}, \bibinfo {author}
		{\bibfnamefont {M.}~\bibnamefont {Schramb{\"{o}}ck}}, \bibinfo {author}
		{\bibfnamefont {G.}~\bibnamefont {Steinhauser}}, \bibinfo {author}
		{\bibfnamefont {H.}~\bibnamefont {Ritsch}}, \bibinfo {author} {\bibfnamefont
			{J.}~\bibnamefont {Schmiedmayer}}, \ and\ \bibinfo {author} {\bibfnamefont
			{J.}~\bibnamefont {Majer}},\ }\href {\doibase 10.1103/PhysRevLett.107.060502}
	{\bibfield  {journal} {\bibinfo  {journal} {Physical Review Letters}\
		}\textbf {\bibinfo {volume} {107}},\ \bibinfo {pages} {060502} (\bibinfo
		{year} {2011})}\BibitemShut {NoStop}%
	\bibitem [{\citenamefont {Ranjan}\ \emph {et~al.}(2013)\citenamefont {Ranjan},
		\citenamefont {{De Lange}}, \citenamefont {Schutjens}, \citenamefont
		{Debelhoir}, \citenamefont {Groen}, \citenamefont {Szombati}, \citenamefont
		{Thoen}, \citenamefont {Klapwijk}, \citenamefont {Hanson},\ and\
		\citenamefont {Dicarlo}}]{Ranjan2013}%
	\BibitemOpen
	\bibfield  {author} {\bibinfo {author} {\bibfnamefont {V.}~\bibnamefont
			{Ranjan}}, \bibinfo {author} {\bibfnamefont {G.}~\bibnamefont {{De Lange}}},
		\bibinfo {author} {\bibfnamefont {R.}~\bibnamefont {Schutjens}}, \bibinfo
		{author} {\bibfnamefont {T.}~\bibnamefont {Debelhoir}}, \bibinfo {author}
		{\bibfnamefont {J.~P.}\ \bibnamefont {Groen}}, \bibinfo {author}
		{\bibfnamefont {D.}~\bibnamefont {Szombati}}, \bibinfo {author}
		{\bibfnamefont {D.~J.}\ \bibnamefont {Thoen}}, \bibinfo {author}
		{\bibfnamefont {T.~M.}\ \bibnamefont {Klapwijk}}, \bibinfo {author}
		{\bibfnamefont {R.}~\bibnamefont {Hanson}}, \ and\ \bibinfo {author}
		{\bibfnamefont {L.}~\bibnamefont {Dicarlo}},\ }\href {\doibase
		10.1103/PhysRevLett.110.067004} {\bibfield  {journal} {\bibinfo  {journal}
			{Physical Review Letters}\ }\textbf {\bibinfo {volume} {110}},\ \bibinfo
		{pages} {067004} (\bibinfo {year} {2013})}\BibitemShut {NoStop}%
	\bibitem [{\citenamefont {Regal}\ \emph {et~al.}(2008)\citenamefont {Regal},
		\citenamefont {Teufel},\ and\ \citenamefont {Lehnert}}]{Regal2008}%
	\BibitemOpen
	\bibfield  {author} {\bibinfo {author} {\bibfnamefont {C.~A.}\ \bibnamefont
			{Regal}}, \bibinfo {author} {\bibfnamefont {J.~D.}\ \bibnamefont {Teufel}}, \
		and\ \bibinfo {author} {\bibfnamefont {K.~W.}\ \bibnamefont {Lehnert}},\
	}\href {\doibase 10.1038/nphys974} {\bibfield  {journal} {\bibinfo  {journal}
			{Nature Physics}\ }\textbf {\bibinfo {volume} {4}},\ \bibinfo {pages} {555}
		(\bibinfo {year} {2008})}\BibitemShut {NoStop}%
	\bibitem [{\citenamefont {Teufel}\ \emph {et~al.}(2008)\citenamefont {Teufel},
		\citenamefont {Regal},\ and\ \citenamefont {Lehnert}}]{Teufel2008}%
	\BibitemOpen
	\bibfield  {author} {\bibinfo {author} {\bibfnamefont {J.~D.}\ \bibnamefont
			{Teufel}}, \bibinfo {author} {\bibfnamefont {C.~A.}\ \bibnamefont {Regal}}, \
		and\ \bibinfo {author} {\bibfnamefont {K.~W.}\ \bibnamefont {Lehnert}},\
	}\href {\doibase 10.1088/1367-2630/10/9/095002} {\bibfield  {journal}
		{\bibinfo  {journal} {New Journal of Physics}\ }\textbf {\bibinfo {volume}
			{10}},\ \bibinfo {pages} {095002} (\bibinfo {year} {2008})}\BibitemShut
	{NoStop}%
	\bibitem [{\citenamefont {Burkard}\ and\ \citenamefont
		{Imamoglu}(2006)}]{Burkard2006}%
	\BibitemOpen
	\bibfield  {author} {\bibinfo {author} {\bibfnamefont {G.}~\bibnamefont
			{Burkard}}\ and\ \bibinfo {author} {\bibfnamefont {A.}~\bibnamefont
			{Imamoglu}},\ }\href {\doibase 10.1103/PhysRevB.74.041307} {\bibfield
		{journal} {\bibinfo  {journal} {Physical Review B}\ }\textbf {\bibinfo
			{volume} {74}},\ \bibinfo {pages} {041307} (\bibinfo {year}
		{2006})}\BibitemShut {NoStop}%
	\bibitem [{\citenamefont {Petersson}\ \emph {et~al.}(2012)\citenamefont
		{Petersson}, \citenamefont {McFaul}, \citenamefont {Schroer}, \citenamefont
		{Jung}, \citenamefont {Taylor}, \citenamefont {Houck},\ and\ \citenamefont
		{Petta}}]{Petersson2012}%
	\BibitemOpen
	\bibfield  {author} {\bibinfo {author} {\bibfnamefont {K.~D.}\ \bibnamefont
			{Petersson}}, \bibinfo {author} {\bibfnamefont {L.~W.}\ \bibnamefont
			{McFaul}}, \bibinfo {author} {\bibfnamefont {M.~D.}\ \bibnamefont {Schroer}},
		\bibinfo {author} {\bibfnamefont {M.}~\bibnamefont {Jung}}, \bibinfo {author}
		{\bibfnamefont {J.~M.}\ \bibnamefont {Taylor}}, \bibinfo {author}
		{\bibfnamefont {A.~A.}\ \bibnamefont {Houck}}, \ and\ \bibinfo {author}
		{\bibfnamefont {J.~R.}\ \bibnamefont {Petta}},\ }\href {\doibase
		10.1038/nature11559} {\bibfield  {journal} {\bibinfo  {journal} {Nature}\
		}\textbf {\bibinfo {volume} {490}},\ \bibinfo {pages} {380} (\bibinfo {year}
		{2012})}\BibitemShut {NoStop}%
	\bibitem [{\citenamefont {Viennot}\ \emph {et~al.}(2015)\citenamefont
		{Viennot}, \citenamefont {Dartiailh}, \citenamefont {Cottet},\ and\
		\citenamefont {Kontos}}]{Viennot2015}%
	\BibitemOpen
	\bibfield  {author} {\bibinfo {author} {\bibfnamefont {J.~J.}\ \bibnamefont
			{Viennot}}, \bibinfo {author} {\bibfnamefont {M.~C.}\ \bibnamefont
			{Dartiailh}}, \bibinfo {author} {\bibfnamefont {A.}~\bibnamefont {Cottet}}, \
		and\ \bibinfo {author} {\bibfnamefont {T.}~\bibnamefont {Kontos}},\ }\href
	{\doibase 10.1126/science.aaa3786} {\bibfield  {journal} {\bibinfo  {journal}
			{Science}\ }\textbf {\bibinfo {volume} {349}},\ \bibinfo {pages} {408}
		(\bibinfo {year} {2015})}\BibitemShut {NoStop}%
	\bibitem [{\citenamefont {Liu}\ \emph {et~al.}(2015)\citenamefont {Liu},
		\citenamefont {Stehlik}, \citenamefont {Eichler}, \citenamefont {Gullans},
		\citenamefont {Taylor},\ and\ \citenamefont {Petta}}]{Liu2015}%
	\BibitemOpen
	\bibfield  {author} {\bibinfo {author} {\bibfnamefont {Y.~Y.}\ \bibnamefont
			{Liu}}, \bibinfo {author} {\bibfnamefont {J.}~\bibnamefont {Stehlik}},
		\bibinfo {author} {\bibfnamefont {C.}~\bibnamefont {Eichler}}, \bibinfo
		{author} {\bibfnamefont {M.~J.}\ \bibnamefont {Gullans}}, \bibinfo {author}
		{\bibfnamefont {J.~M.}\ \bibnamefont {Taylor}}, \ and\ \bibinfo {author}
		{\bibfnamefont {J.~R.}\ \bibnamefont {Petta}},\ }\href@noop {} {\bibfield
		{journal} {\bibinfo  {journal} {Science}\ }\textbf {\bibinfo {volume}
			{347}},\ \bibinfo {pages} {285} (\bibinfo {year} {2015})}\BibitemShut
	{NoStop}%
	\bibitem [{\citenamefont {{De Lange}}\ \emph {et~al.}(2015)\citenamefont {{De
				Lange}}, \citenamefont {{Van Heck}}, \citenamefont {Bruno}, \citenamefont
		{{Van Woerkom}}, \citenamefont {Geresdi}, \citenamefont {Plissard},
		\citenamefont {Bakkers}, \citenamefont {Akhmerov},\ and\ \citenamefont
		{DiCarlo}}]{DeLange2015}%
	\BibitemOpen
	\bibfield  {author} {\bibinfo {author} {\bibfnamefont {G.}~\bibnamefont {{De
					Lange}}}, \bibinfo {author} {\bibfnamefont {B.}~\bibnamefont {{Van Heck}}},
		\bibinfo {author} {\bibfnamefont {A.}~\bibnamefont {Bruno}}, \bibinfo
		{author} {\bibfnamefont {D.~J.}\ \bibnamefont {{Van Woerkom}}}, \bibinfo
		{author} {\bibfnamefont {A.}~\bibnamefont {Geresdi}}, \bibinfo {author}
		{\bibfnamefont {S.~R.}\ \bibnamefont {Plissard}}, \bibinfo {author}
		{\bibfnamefont {E.~P. A.~M.}\ \bibnamefont {Bakkers}}, \bibinfo {author}
		{\bibfnamefont {A.~R.}\ \bibnamefont {Akhmerov}}, \ and\ \bibinfo {author}
		{\bibfnamefont {L.}~\bibnamefont {DiCarlo}},\ }\href {\doibase
		10.1103/PhysRevLett.115.127002} {\bibfield  {journal} {\bibinfo  {journal}
			{Physical Review Letters}\ }\textbf {\bibinfo {volume} {115}},\ \bibinfo
		{pages} {127002} (\bibinfo {year} {2015})}\BibitemShut {NoStop}%
	\bibitem [{\citenamefont {Larsen}\ \emph {et~al.}(2015)\citenamefont {Larsen},
		\citenamefont {Petersson}, \citenamefont {Kuemmeth}, \citenamefont
		{Jespersen}, \citenamefont {Krogstrup}, \citenamefont {Nyg{\aa}rd},\ and\
		\citenamefont {Marcus}}]{Larsen2015}%
	\BibitemOpen
	\bibfield  {author} {\bibinfo {author} {\bibfnamefont {T.~W.}\ \bibnamefont
			{Larsen}}, \bibinfo {author} {\bibfnamefont {K.~D.}\ \bibnamefont
			{Petersson}}, \bibinfo {author} {\bibfnamefont {F.}~\bibnamefont {Kuemmeth}},
		\bibinfo {author} {\bibfnamefont {T.~S.}\ \bibnamefont {Jespersen}}, \bibinfo
		{author} {\bibfnamefont {P.}~\bibnamefont {Krogstrup}}, \bibinfo {author}
		{\bibfnamefont {J.}~\bibnamefont {Nyg{\aa}rd}}, \ and\ \bibinfo {author}
		{\bibfnamefont {C.~M.}\ \bibnamefont {Marcus}},\ }\href {\doibase
		10.1103/PhysRevLett.115.127001} {\bibfield  {journal} {\bibinfo  {journal}
			{Physical Review Letters}\ }\textbf {\bibinfo {volume} {115}},\ \bibinfo
		{pages} {127001} (\bibinfo {year} {2015})}\BibitemShut {NoStop}%
	\bibitem [{\citenamefont {Kroll}\ \emph {et~al.}(2018)\citenamefont {Kroll},
		\citenamefont {Uilhoorn}, \citenamefont {van~der Enden}, \citenamefont
		{de~Jong}, \citenamefont {Watanabe}, \citenamefont {Taniguchi}, \citenamefont
		{Goswami}, \citenamefont {Cassidy},\ and\ \citenamefont
		{Kouwenhoven}}]{Kroll2018}%
	\BibitemOpen
	\bibfield  {author} {\bibinfo {author} {\bibfnamefont {J.~G.}\ \bibnamefont
			{Kroll}}, \bibinfo {author} {\bibfnamefont {W.}~\bibnamefont {Uilhoorn}},
		\bibinfo {author} {\bibfnamefont {K.~L.}\ \bibnamefont {van~der Enden}},
		\bibinfo {author} {\bibfnamefont {D.}~\bibnamefont {de~Jong}}, \bibinfo
		{author} {\bibfnamefont {K.}~\bibnamefont {Watanabe}}, \bibinfo {author}
		{\bibfnamefont {T.}~\bibnamefont {Taniguchi}}, \bibinfo {author}
		{\bibfnamefont {S.}~\bibnamefont {Goswami}}, \bibinfo {author} {\bibfnamefont
			{M.~C.}\ \bibnamefont {Cassidy}}, \ and\ \bibinfo {author} {\bibfnamefont
			{L.~P.}\ \bibnamefont {Kouwenhoven}},\ }\href@noop {} {\  (\bibinfo {year}
		{2018})},\ \Eprint {http://arxiv.org/abs/1806.10534} {arXiv:1806.10534}
	\BibitemShut {NoStop}%
	\bibitem [{\citenamefont {Blais}\ \emph {et~al.}(2004)\citenamefont {Blais},
		\citenamefont {Huang}, \citenamefont {Wallraff}, \citenamefont {Girvin},\
		and\ \citenamefont {Schoelkopf}}]{Blais2004}%
	\BibitemOpen
	\bibfield  {author} {\bibinfo {author} {\bibfnamefont {A.}~\bibnamefont
			{Blais}}, \bibinfo {author} {\bibfnamefont {R.~S.}\ \bibnamefont {Huang}},
		\bibinfo {author} {\bibfnamefont {A.}~\bibnamefont {Wallraff}}, \bibinfo
		{author} {\bibfnamefont {S.~M.}\ \bibnamefont {Girvin}}, \ and\ \bibinfo
		{author} {\bibfnamefont {R.~J.}\ \bibnamefont {Schoelkopf}},\ }\href
	{\doibase 10.1103/PhysRevA.69.062320} {\bibfield  {journal} {\bibinfo
			{journal} {Physical Review A}\ }\textbf {\bibinfo {volume} {69}},\ \bibinfo
		{pages} {062320} (\bibinfo {year} {2004})}\BibitemShut {NoStop}%
	\bibitem [{\citenamefont {Wallraff}\ \emph {et~al.}(2004)\citenamefont
		{Wallraff}, \citenamefont {Schuster}, \citenamefont {Blais}, \citenamefont
		{Frunzio}, \citenamefont {Huang}, \citenamefont {Majer}, \citenamefont
		{Kumar}, \citenamefont {Girvin},\ and\ \citenamefont
		{Schoelkopf}}]{Wallraff2004}%
	\BibitemOpen
	\bibfield  {author} {\bibinfo {author} {\bibfnamefont {A.}~\bibnamefont
			{Wallraff}}, \bibinfo {author} {\bibfnamefont {D.~I.}\ \bibnamefont
			{Schuster}}, \bibinfo {author} {\bibfnamefont {A.}~\bibnamefont {Blais}},
		\bibinfo {author} {\bibfnamefont {L.}~\bibnamefont {Frunzio}}, \bibinfo
		{author} {\bibfnamefont {R.~S.}\ \bibnamefont {Huang}}, \bibinfo {author}
		{\bibfnamefont {J.}~\bibnamefont {Majer}}, \bibinfo {author} {\bibfnamefont
			{S.}~\bibnamefont {Kumar}}, \bibinfo {author} {\bibfnamefont {S.~M.}\
			\bibnamefont {Girvin}}, \ and\ \bibinfo {author} {\bibfnamefont {R.~J.}\
			\bibnamefont {Schoelkopf}},\ }\href {\doibase 10.1038/nature02851} {\bibfield
		{journal} {\bibinfo  {journal} {Nature}\ }\textbf {\bibinfo {volume}
			{431}},\ \bibinfo {pages} {162} (\bibinfo {year} {2004})}\BibitemShut
	{NoStop}%
	\bibitem [{\citenamefont {Wallraff}\ \emph {et~al.}(2005)\citenamefont
		{Wallraff}, \citenamefont {Schuster}, \citenamefont {Blais}, \citenamefont
		{Frunzio}, \citenamefont {Majer}, \citenamefont {Devoret}, \citenamefont
		{Girvin},\ and\ \citenamefont {Schoelkopf}}]{Wallraff2005}%
	\BibitemOpen
	\bibfield  {author} {\bibinfo {author} {\bibfnamefont {A.}~\bibnamefont
			{Wallraff}}, \bibinfo {author} {\bibfnamefont {D.~I.}\ \bibnamefont
			{Schuster}}, \bibinfo {author} {\bibfnamefont {A.}~\bibnamefont {Blais}},
		\bibinfo {author} {\bibfnamefont {L.}~\bibnamefont {Frunzio}}, \bibinfo
		{author} {\bibfnamefont {J.}~\bibnamefont {Majer}}, \bibinfo {author}
		{\bibfnamefont {M.~H.}\ \bibnamefont {Devoret}}, \bibinfo {author}
		{\bibfnamefont {S.~M.}\ \bibnamefont {Girvin}}, \ and\ \bibinfo {author}
		{\bibfnamefont {R.~J.}\ \bibnamefont {Schoelkopf}},\ }\href {\doibase
		10.1103/PhysRevLett.95.060501} {\bibfield  {journal} {\bibinfo  {journal}
			{Physical Review Letters}\ }\textbf {\bibinfo {volume} {95}},\ \bibinfo
		{pages} {060501} (\bibinfo {year} {2005})}\BibitemShut {NoStop}%
	\bibitem [{\citenamefont {Majer}\ \emph {et~al.}(2007)\citenamefont {Majer},
		\citenamefont {Chow}, \citenamefont {Gambetta}, \citenamefont {Koch},
		\citenamefont {Johnson}, \citenamefont {Schreier}, \citenamefont {Frunzio},
		\citenamefont {Schuster}, \citenamefont {Houck}, \citenamefont {Wallraff},
		\citenamefont {Blais}, \citenamefont {Devoret}, \citenamefont {Girvin},\ and\
		\citenamefont {Schoelkopf}}]{Majer2007}%
	\BibitemOpen
	\bibfield  {author} {\bibinfo {author} {\bibfnamefont {J.}~\bibnamefont
			{Majer}}, \bibinfo {author} {\bibfnamefont {J.~M.}\ \bibnamefont {Chow}},
		\bibinfo {author} {\bibfnamefont {J.~M.}\ \bibnamefont {Gambetta}}, \bibinfo
		{author} {\bibfnamefont {J.}~\bibnamefont {Koch}}, \bibinfo {author}
		{\bibfnamefont {B.~R.}\ \bibnamefont {Johnson}}, \bibinfo {author}
		{\bibfnamefont {J.~A.}\ \bibnamefont {Schreier}}, \bibinfo {author}
		{\bibfnamefont {L.}~\bibnamefont {Frunzio}}, \bibinfo {author} {\bibfnamefont
			{D.~I.}\ \bibnamefont {Schuster}}, \bibinfo {author} {\bibfnamefont {A.~A.}\
			\bibnamefont {Houck}}, \bibinfo {author} {\bibfnamefont {A.}~\bibnamefont
			{Wallraff}}, \bibinfo {author} {\bibfnamefont {A.}~\bibnamefont {Blais}},
		\bibinfo {author} {\bibfnamefont {M.~H.}\ \bibnamefont {Devoret}}, \bibinfo
		{author} {\bibfnamefont {S.~M.}\ \bibnamefont {Girvin}}, \ and\ \bibinfo
		{author} {\bibfnamefont {R.~J.}\ \bibnamefont {Schoelkopf}},\ }\href
	{\doibase 10.1038/nature06184} {\bibfield  {journal} {\bibinfo  {journal}
			{Nature}\ }\textbf {\bibinfo {volume} {449}},\ \bibinfo {pages} {443}
		(\bibinfo {year} {2007})}\BibitemShut {NoStop}%
	\bibitem [{\citenamefont {Stockklauser}\ \emph {et~al.}(2017)\citenamefont
		{Stockklauser}, \citenamefont {Scarlino}, \citenamefont {Koski},
		\citenamefont {Gasparinetti}, \citenamefont {Andersen}, \citenamefont
		{Reichl}, \citenamefont {Wegscheider}, \citenamefont {Ihn}, \citenamefont
		{Ensslin},\ and\ \citenamefont {Wallraff}}]{Stockklauser2017}%
	\BibitemOpen
	\bibfield  {author} {\bibinfo {author} {\bibfnamefont {A.}~\bibnamefont
			{Stockklauser}}, \bibinfo {author} {\bibfnamefont {P.}~\bibnamefont
			{Scarlino}}, \bibinfo {author} {\bibfnamefont {J.~V.}\ \bibnamefont {Koski}},
		\bibinfo {author} {\bibfnamefont {S.}~\bibnamefont {Gasparinetti}}, \bibinfo
		{author} {\bibfnamefont {C.~K.}\ \bibnamefont {Andersen}}, \bibinfo {author}
		{\bibfnamefont {C.}~\bibnamefont {Reichl}}, \bibinfo {author} {\bibfnamefont
			{W.}~\bibnamefont {Wegscheider}}, \bibinfo {author} {\bibfnamefont
			{T.}~\bibnamefont {Ihn}}, \bibinfo {author} {\bibfnamefont {K.}~\bibnamefont
			{Ensslin}}, \ and\ \bibinfo {author} {\bibfnamefont {A.}~\bibnamefont
			{Wallraff}},\ }\href {\doibase 10.1103/PhysRevX.7.011030} {\bibfield
		{journal} {\bibinfo  {journal} {Physical Review X}\ }\textbf {\bibinfo
			{volume} {7}},\ \bibinfo {pages} {011030} (\bibinfo {year}
		{2017})}\BibitemShut {NoStop}%
	\bibitem [{\citenamefont {Landig}\ \emph {et~al.}(2017)\citenamefont {Landig},
		\citenamefont {Koski}, \citenamefont {Scarlino}, \citenamefont {Mendes},
		\citenamefont {Blais}, \citenamefont {Reichl}, \citenamefont {Wegscheider},
		\citenamefont {Wallraff}, \citenamefont {Ensslin},\ and\ \citenamefont
		{Ihn}}]{Landig2018}%
	\BibitemOpen
	\bibfield  {author} {\bibinfo {author} {\bibfnamefont {A.~J.}\ \bibnamefont
			{Landig}}, \bibinfo {author} {\bibfnamefont {J.~V.}\ \bibnamefont {Koski}},
		\bibinfo {author} {\bibfnamefont {P.}~\bibnamefont {Scarlino}}, \bibinfo
		{author} {\bibfnamefont {U.~C.}\ \bibnamefont {Mendes}}, \bibinfo {author}
		{\bibfnamefont {A.}~\bibnamefont {Blais}}, \bibinfo {author} {\bibfnamefont
			{C.}~\bibnamefont {Reichl}}, \bibinfo {author} {\bibfnamefont
			{W.}~\bibnamefont {Wegscheider}}, \bibinfo {author} {\bibfnamefont
			{A.}~\bibnamefont {Wallraff}}, \bibinfo {author} {\bibfnamefont
			{K.}~\bibnamefont {Ensslin}}, \ and\ \bibinfo {author} {\bibfnamefont
			{T.}~\bibnamefont {Ihn}},\ }\href {\doibase 10.1038/s41586-018-0365-y}
	{\bibfield  {journal} {\bibinfo  {journal} {Nature}\ }\textbf {\bibinfo
			{volume} {560}},\ \bibinfo {pages} {179} (\bibinfo {year}
		{2017})}\BibitemShut {NoStop}%
	\bibitem [{\citenamefont {Mi}\ \emph {et~al.}(2018)\citenamefont {Mi},
		\citenamefont {Benito}, \citenamefont {Putz}, \citenamefont {Zajac},
		\citenamefont {Taylor}, \citenamefont {Burkard},\ and\ \citenamefont
		{Petta}}]{Mi2018}%
	\BibitemOpen
	\bibfield  {author} {\bibinfo {author} {\bibfnamefont {X.}~\bibnamefont
			{Mi}}, \bibinfo {author} {\bibfnamefont {M.}~\bibnamefont {Benito}}, \bibinfo
		{author} {\bibfnamefont {S.}~\bibnamefont {Putz}}, \bibinfo {author}
		{\bibfnamefont {D.~M.}\ \bibnamefont {Zajac}}, \bibinfo {author}
		{\bibfnamefont {J.~M.}\ \bibnamefont {Taylor}}, \bibinfo {author}
		{\bibfnamefont {G.}~\bibnamefont {Burkard}}, \ and\ \bibinfo {author}
		{\bibfnamefont {J.~R.}\ \bibnamefont {Petta}},\ }\href {\doibase
		10.1038/nature25769} {\bibfield  {journal} {\bibinfo  {journal} {Nature}\
		}\textbf {\bibinfo {volume} {555}},\ \bibinfo {pages} {599} (\bibinfo {year}
		{2018})}\BibitemShut {NoStop}%
	\bibitem [{\citenamefont {Samkharadze}\ \emph {et~al.}(2018)\citenamefont
		{Samkharadze}, \citenamefont {Zheng}, \citenamefont {Kalhor}, \citenamefont
		{Brousse}, \citenamefont {Sammak}, \citenamefont {Mendes}, \citenamefont
		{Blais}, \citenamefont {Scappucci},\ and\ \citenamefont
		{Vandersypen}}]{Samkharadze2018}%
	\BibitemOpen
	\bibfield  {author} {\bibinfo {author} {\bibfnamefont {N.}~\bibnamefont
			{Samkharadze}}, \bibinfo {author} {\bibfnamefont {G.}~\bibnamefont {Zheng}},
		\bibinfo {author} {\bibfnamefont {N.}~\bibnamefont {Kalhor}}, \bibinfo
		{author} {\bibfnamefont {D.}~\bibnamefont {Brousse}}, \bibinfo {author}
		{\bibfnamefont {A.}~\bibnamefont {Sammak}}, \bibinfo {author} {\bibfnamefont
			{U.~C.}\ \bibnamefont {Mendes}}, \bibinfo {author} {\bibfnamefont
			{A.}~\bibnamefont {Blais}}, \bibinfo {author} {\bibfnamefont
			{G.}~\bibnamefont {Scappucci}}, \ and\ \bibinfo {author} {\bibfnamefont
			{L.~M.~K.}\ \bibnamefont {Vandersypen}},\ }\href {\doibase
		10.1126/science.aar4054} {\bibfield  {journal} {\bibinfo  {journal}
			{Science}\ }\textbf {\bibinfo {volume} {359}},\ \bibinfo {pages} {1123}
		(\bibinfo {year} {2018})}\BibitemShut {NoStop}%
	\bibitem [{\citenamefont {Hyart}\ \emph {et~al.}(2013)\citenamefont {Hyart},
		\citenamefont {{Van Heck}}, \citenamefont {Fulga}, \citenamefont {Burrello},
		\citenamefont {Akhmerov},\ and\ \citenamefont {Beenakker}}]{Hyart2013}%
	\BibitemOpen
	\bibfield  {author} {\bibinfo {author} {\bibfnamefont {T.}~\bibnamefont
			{Hyart}}, \bibinfo {author} {\bibfnamefont {B.}~\bibnamefont {{Van Heck}}},
		\bibinfo {author} {\bibfnamefont {I.~C.}\ \bibnamefont {Fulga}}, \bibinfo
		{author} {\bibfnamefont {M.}~\bibnamefont {Burrello}}, \bibinfo {author}
		{\bibfnamefont {A.~R.}\ \bibnamefont {Akhmerov}}, \ and\ \bibinfo {author}
		{\bibfnamefont {C.~W.~J.}\ \bibnamefont {Beenakker}},\ }\href {\doibase
		10.1103/PhysRevB.88.035121} {\bibfield  {journal} {\bibinfo  {journal}
			{Physical Review B}\ }\textbf {\bibinfo {volume} {88}},\ \bibinfo {pages}
		{035121} (\bibinfo {year} {2013})}\BibitemShut {NoStop}%
	\bibitem [{\citenamefont {Plugge}\ \emph {et~al.}(2017)\citenamefont {Plugge},
		\citenamefont {Rasmussen}, \citenamefont {Egger},\ and\ \citenamefont
		{Flensberg}}]{Plugge2017}%
	\BibitemOpen
	\bibfield  {author} {\bibinfo {author} {\bibfnamefont {S.}~\bibnamefont
			{Plugge}}, \bibinfo {author} {\bibfnamefont {A.}~\bibnamefont {Rasmussen}},
		\bibinfo {author} {\bibfnamefont {R.}~\bibnamefont {Egger}}, \ and\ \bibinfo
		{author} {\bibfnamefont {K.}~\bibnamefont {Flensberg}},\ }\href {\doibase
		10.1088/1367-2630/aa54e1} {\bibfield  {journal} {\bibinfo  {journal} {New
				Journal of Physics}\ }\textbf {\bibinfo {volume} {19}},\ \bibinfo {pages}
		{012001} (\bibinfo {year} {2017})}\BibitemShut {NoStop}%
	\bibitem [{\citenamefont {Koch}\ \emph {et~al.}(2007)\citenamefont {Koch},
		\citenamefont {Yu}, \citenamefont {Gambetta}, \citenamefont {Houck},
		\citenamefont {Schuster}, \citenamefont {Majer}, \citenamefont {Blais},
		\citenamefont {Devoret}, \citenamefont {Girvin},\ and\ \citenamefont
		{Schoelkopf}}]{Koch2007}%
	\BibitemOpen
	\bibfield  {author} {\bibinfo {author} {\bibfnamefont {J.}~\bibnamefont
			{Koch}}, \bibinfo {author} {\bibfnamefont {T.}~\bibnamefont {Yu}}, \bibinfo
		{author} {\bibfnamefont {J.}~\bibnamefont {Gambetta}}, \bibinfo {author}
		{\bibfnamefont {A.~A.}\ \bibnamefont {Houck}}, \bibinfo {author}
		{\bibfnamefont {D.~I.}\ \bibnamefont {Schuster}}, \bibinfo {author}
		{\bibfnamefont {J.}~\bibnamefont {Majer}}, \bibinfo {author} {\bibfnamefont
			{A.}~\bibnamefont {Blais}}, \bibinfo {author} {\bibfnamefont {M.~H.}\
			\bibnamefont {Devoret}}, \bibinfo {author} {\bibfnamefont {S.~M.}\
			\bibnamefont {Girvin}}, \ and\ \bibinfo {author} {\bibfnamefont {R.~J.}\
			\bibnamefont {Schoelkopf}},\ }\href {\doibase 10.1103/PhysRevA.76.042319}
	{\bibfield  {journal} {\bibinfo  {journal} {Physical Review A}\ }\textbf
		{\bibinfo {volume} {76}},\ \bibinfo {pages} {042319} (\bibinfo {year}
		{2007})}\BibitemShut {NoStop}%
	\bibitem [{\citenamefont {Harvey}\ \emph {et~al.}(2018)\citenamefont {Harvey},
		\citenamefont {B{\o}ttcher}, \citenamefont {Orona}, \citenamefont {Bartlett},
		\citenamefont {Doherty},\ and\ \citenamefont {Yacoby}}]{Harvey}%
	\BibitemOpen
	\bibfield  {author} {\bibinfo {author} {\bibfnamefont {S.~P.}\ \bibnamefont
			{Harvey}}, \bibinfo {author} {\bibfnamefont {C.~G.~L.}\ \bibnamefont
			{B{\o}ttcher}}, \bibinfo {author} {\bibfnamefont {L.~A.}\ \bibnamefont
			{Orona}}, \bibinfo {author} {\bibfnamefont {S.~D.}\ \bibnamefont {Bartlett}},
		\bibinfo {author} {\bibfnamefont {A.~C.}\ \bibnamefont {Doherty}}, \ and\
		\bibinfo {author} {\bibfnamefont {A.}~\bibnamefont {Yacoby}},\ }\href@noop {}
	{\  (\bibinfo {year} {2018})},\ \Eprint {http://arxiv.org/abs/1801.04858}
	{arXiv:1801.04858} \BibitemShut {NoStop}%
	\bibitem [{\citenamefont {Gao}\ \emph {et~al.}(2008)\citenamefont {Gao},
		\citenamefont {Daal}, \citenamefont {Vayonakis}, \citenamefont {Kumar},
		\citenamefont {Zmuidzinas}, \citenamefont {Sadoulet}, \citenamefont {Mazin},
		\citenamefont {Day},\ and\ \citenamefont {Leduc}}]{Gao2008}%
	\BibitemOpen
	\bibfield  {author} {\bibinfo {author} {\bibfnamefont {J.}~\bibnamefont
			{Gao}}, \bibinfo {author} {\bibfnamefont {M.}~\bibnamefont {Daal}}, \bibinfo
		{author} {\bibfnamefont {A.}~\bibnamefont {Vayonakis}}, \bibinfo {author}
		{\bibfnamefont {S.}~\bibnamefont {Kumar}}, \bibinfo {author} {\bibfnamefont
			{J.}~\bibnamefont {Zmuidzinas}}, \bibinfo {author} {\bibfnamefont
			{B.}~\bibnamefont {Sadoulet}}, \bibinfo {author} {\bibfnamefont {B.~A.}\
			\bibnamefont {Mazin}}, \bibinfo {author} {\bibfnamefont {P.~K.}\ \bibnamefont
			{Day}}, \ and\ \bibinfo {author} {\bibfnamefont {H.~G.}\ \bibnamefont
			{Leduc}},\ }\href {\doibase 10.1063/1.2906373} {\bibfield  {journal}
		{\bibinfo  {journal} {Applied Physics Letters}\ }\textbf {\bibinfo {volume}
			{92}},\ \bibinfo {pages} {152505} (\bibinfo {year} {2008})}\BibitemShut
	{NoStop}%
	\bibitem [{\citenamefont {Barends}\ \emph {et~al.}(2011)\citenamefont
		{Barends}, \citenamefont {Wenner}, \citenamefont {Lenander}, \citenamefont
		{Chen}, \citenamefont {Bialczak}, \citenamefont {Kelly}, \citenamefont
		{Lucero}, \citenamefont {O'Malley}, \citenamefont {Mariantoni}, \citenamefont
		{Sank}, \citenamefont {Wang}, \citenamefont {White}, \citenamefont {Yin},
		\citenamefont {Zhao}, \citenamefont {Cleland}, \citenamefont {Martinis},\
		and\ \citenamefont {Baselmans}}]{Barends2011}%
	\BibitemOpen
	\bibfield  {author} {\bibinfo {author} {\bibfnamefont {R.}~\bibnamefont
			{Barends}}, \bibinfo {author} {\bibfnamefont {J.}~\bibnamefont {Wenner}},
		\bibinfo {author} {\bibfnamefont {M.}~\bibnamefont {Lenander}}, \bibinfo
		{author} {\bibfnamefont {Y.}~\bibnamefont {Chen}}, \bibinfo {author}
		{\bibfnamefont {R.~C.}\ \bibnamefont {Bialczak}}, \bibinfo {author}
		{\bibfnamefont {J.}~\bibnamefont {Kelly}}, \bibinfo {author} {\bibfnamefont
			{E.}~\bibnamefont {Lucero}}, \bibinfo {author} {\bibfnamefont
			{P.}~\bibnamefont {O'Malley}}, \bibinfo {author} {\bibfnamefont
			{M.}~\bibnamefont {Mariantoni}}, \bibinfo {author} {\bibfnamefont
			{D.}~\bibnamefont {Sank}}, \bibinfo {author} {\bibfnamefont {H.}~\bibnamefont
			{Wang}}, \bibinfo {author} {\bibfnamefont {T.~C.}\ \bibnamefont {White}},
		\bibinfo {author} {\bibfnamefont {Y.}~\bibnamefont {Yin}}, \bibinfo {author}
		{\bibfnamefont {J.}~\bibnamefont {Zhao}}, \bibinfo {author} {\bibfnamefont
			{A.~N.}\ \bibnamefont {Cleland}}, \bibinfo {author} {\bibfnamefont {J.~M.}\
			\bibnamefont {Martinis}}, \ and\ \bibinfo {author} {\bibfnamefont {J.~J.~A.}\
			\bibnamefont {Baselmans}},\ }\href {\doibase 10.1063/1.3638063} {\bibfield
		{journal} {\bibinfo  {journal} {Applied Physics Letters}\ }\textbf {\bibinfo
			{volume} {99}},\ \bibinfo {pages} {113507} (\bibinfo {year}
		{2011})}\BibitemShut {NoStop}%
	\bibitem [{\citenamefont {Song}(2011)}]{Song2011}%
	\BibitemOpen
	\bibfield  {author} {\bibinfo {author} {\bibfnamefont {C.}~\bibnamefont
			{Song}},\ }\emph {\bibinfo {title} {Dissertation}},\ \href@noop {} {Ph.D.
		thesis} (\bibinfo {year} {2011})\BibitemShut {NoStop}%
	\bibitem [{\citenamefont {Megrant}\ \emph {et~al.}(2012)\citenamefont
		{Megrant}, \citenamefont {Neill}, \citenamefont {Barends}, \citenamefont
		{Chiaro}, \citenamefont {Chen}, \citenamefont {Feigl}, \citenamefont {Kelly},
		\citenamefont {Lucero}, \citenamefont {Mariantoni}, \citenamefont {O'Malley},
		\citenamefont {Sank}, \citenamefont {Vainsencher}, \citenamefont {Wenner},
		\citenamefont {White}, \citenamefont {Yin}, \citenamefont {Zhao},
		\citenamefont {Palmstr{\o}m}, \citenamefont {Martinis},\ and\ \citenamefont
		{Cleland}}]{Megrant2012}%
	\BibitemOpen
	\bibfield  {author} {\bibinfo {author} {\bibfnamefont {A.}~\bibnamefont
			{Megrant}}, \bibinfo {author} {\bibfnamefont {C.}~\bibnamefont {Neill}},
		\bibinfo {author} {\bibfnamefont {R.}~\bibnamefont {Barends}}, \bibinfo
		{author} {\bibfnamefont {B.}~\bibnamefont {Chiaro}}, \bibinfo {author}
		{\bibfnamefont {Y.}~\bibnamefont {Chen}}, \bibinfo {author} {\bibfnamefont
			{L.}~\bibnamefont {Feigl}}, \bibinfo {author} {\bibfnamefont
			{J.}~\bibnamefont {Kelly}}, \bibinfo {author} {\bibfnamefont
			{E.}~\bibnamefont {Lucero}}, \bibinfo {author} {\bibfnamefont
			{M.}~\bibnamefont {Mariantoni}}, \bibinfo {author} {\bibfnamefont {P.~J.~J.}\
			\bibnamefont {O'Malley}}, \bibinfo {author} {\bibfnamefont {D.}~\bibnamefont
			{Sank}}, \bibinfo {author} {\bibfnamefont {A.}~\bibnamefont {Vainsencher}},
		\bibinfo {author} {\bibfnamefont {J.}~\bibnamefont {Wenner}}, \bibinfo
		{author} {\bibfnamefont {T.~C.}\ \bibnamefont {White}}, \bibinfo {author}
		{\bibfnamefont {Y.}~\bibnamefont {Yin}}, \bibinfo {author} {\bibfnamefont
			{J.}~\bibnamefont {Zhao}}, \bibinfo {author} {\bibfnamefont {C.~J.}\
			\bibnamefont {Palmstr{\o}m}}, \bibinfo {author} {\bibfnamefont {J.~M.}\
			\bibnamefont {Martinis}}, \ and\ \bibinfo {author} {\bibfnamefont {A.~N.}\
			\bibnamefont {Cleland}},\ }\href {\doibase 10.1063/1.3693409} {\bibfield
		{journal} {\bibinfo  {journal} {Applied Physics Letters}\ }\textbf {\bibinfo
			{volume} {100}},\ \bibinfo {pages} {113510} (\bibinfo {year}
		{2012})}\BibitemShut {NoStop}%
	\bibitem [{\citenamefont {Bruno}\ \emph {et~al.}(2015)\citenamefont {Bruno},
		\citenamefont {{De Lange}}, \citenamefont {Asaad}, \citenamefont {{Van Der
				Enden}}, \citenamefont {Langford},\ and\ \citenamefont
		{Dicarlo}}]{Bruno2015}%
	\BibitemOpen
	\bibfield  {author} {\bibinfo {author} {\bibfnamefont {A.}~\bibnamefont
			{Bruno}}, \bibinfo {author} {\bibfnamefont {G.}~\bibnamefont {{De Lange}}},
		\bibinfo {author} {\bibfnamefont {S.}~\bibnamefont {Asaad}}, \bibinfo
		{author} {\bibfnamefont {K.~L.}\ \bibnamefont {{Van Der Enden}}}, \bibinfo
		{author} {\bibfnamefont {N.~K.}\ \bibnamefont {Langford}}, \ and\ \bibinfo
		{author} {\bibfnamefont {L.}~\bibnamefont {Dicarlo}},\ }\href {\doibase
		10.1063/1.4919761} {\bibfield  {journal} {\bibinfo  {journal} {Applied
				Physics Letters}\ }\textbf {\bibinfo {volume} {106}},\ \bibinfo {pages}
		{182601} (\bibinfo {year} {2015})}\BibitemShut {NoStop}%
	\bibitem [{\citenamefont {Calado}\ \emph {et~al.}(2015)\citenamefont {Calado},
		\citenamefont {Goswami}, \citenamefont {Nanda}, \citenamefont {Diez},
		\citenamefont {Akhmerov}, \citenamefont {Watanabe}, \citenamefont
		{Taniguchi}, \citenamefont {Klapwijk},\ and\ \citenamefont
		{Vandersypen}}]{Calado2015}%
	\BibitemOpen
	\bibfield  {author} {\bibinfo {author} {\bibfnamefont {V.~E.}\ \bibnamefont
			{Calado}}, \bibinfo {author} {\bibfnamefont {S.}~\bibnamefont {Goswami}},
		\bibinfo {author} {\bibfnamefont {G.}~\bibnamefont {Nanda}}, \bibinfo
		{author} {\bibfnamefont {M.}~\bibnamefont {Diez}}, \bibinfo {author}
		{\bibfnamefont {A.~R.}\ \bibnamefont {Akhmerov}}, \bibinfo {author}
		{\bibfnamefont {K.}~\bibnamefont {Watanabe}}, \bibinfo {author}
		{\bibfnamefont {T.}~\bibnamefont {Taniguchi}}, \bibinfo {author}
		{\bibfnamefont {T.~M.}\ \bibnamefont {Klapwijk}}, \ and\ \bibinfo {author}
		{\bibfnamefont {L.~M.~K.}\ \bibnamefont {Vandersypen}},\ }\href {\doibase
		10.1038/nnano.2015.156} {\bibfield  {journal} {\bibinfo  {journal} {Nature
				Nanotechnology}\ }\textbf {\bibinfo {volume} {10}},\ \bibinfo {pages} {761}
		(\bibinfo {year} {2015})}\BibitemShut {NoStop}%
	\bibitem [{\citenamefont {{Van Woerkom}}\ \emph {et~al.}(2015)\citenamefont
		{{Van Woerkom}}, \citenamefont {Geresdi},\ and\ \citenamefont
		{Kouwenhoven}}]{VanWoerkom2015}%
	\BibitemOpen
	\bibfield  {author} {\bibinfo {author} {\bibfnamefont {D.~J.}\ \bibnamefont
			{{Van Woerkom}}}, \bibinfo {author} {\bibfnamefont {A.}~\bibnamefont
			{Geresdi}}, \ and\ \bibinfo {author} {\bibfnamefont {L.~P.}\ \bibnamefont
			{Kouwenhoven}},\ }\href {\doibase 10.1038/nphys3342} {\bibfield  {journal}
		{\bibinfo  {journal} {Nature Physics}\ }\textbf {\bibinfo {volume} {11}},\
		\bibinfo {pages} {547} (\bibinfo {year} {2015})}\BibitemShut {NoStop}%
	\bibitem [{\citenamefont {Vissers}\ \emph {et~al.}(2010)\citenamefont
		{Vissers}, \citenamefont {Gao}, \citenamefont {Wisbey}, \citenamefont {Hite},
		\citenamefont {Tsuei}, \citenamefont {Corcoles}, \citenamefont {Steffen},\
		and\ \citenamefont {Pappas}}]{Vissers2010}%
	\BibitemOpen
	\bibfield  {author} {\bibinfo {author} {\bibfnamefont {M.~R.}\ \bibnamefont
			{Vissers}}, \bibinfo {author} {\bibfnamefont {J.}~\bibnamefont {Gao}},
		\bibinfo {author} {\bibfnamefont {D.~S.}\ \bibnamefont {Wisbey}}, \bibinfo
		{author} {\bibfnamefont {D.~A.}\ \bibnamefont {Hite}}, \bibinfo {author}
		{\bibfnamefont {C.~C.}\ \bibnamefont {Tsuei}}, \bibinfo {author}
		{\bibfnamefont {A.~D.}\ \bibnamefont {Corcoles}}, \bibinfo {author}
		{\bibfnamefont {M.}~\bibnamefont {Steffen}}, \ and\ \bibinfo {author}
		{\bibfnamefont {D.~P.}\ \bibnamefont {Pappas}},\ }\href {\doibase
		10.1063/1.3517252} {\bibfield  {journal} {\bibinfo  {journal} {Applied
				Physics Letters}\ }\textbf {\bibinfo {volume} {97}},\ \bibinfo {pages}
		{232509} (\bibinfo {year} {2010})}\BibitemShut {NoStop}%
	\bibitem [{\citenamefont {Singh}\ \emph {et~al.}(2014)\citenamefont {Singh},
		\citenamefont {Schneider}, \citenamefont {Bosman}, \citenamefont {Merkx},\
		and\ \citenamefont {Steele}}]{Singh2014}%
	\BibitemOpen
	\bibfield  {author} {\bibinfo {author} {\bibfnamefont {V.}~\bibnamefont
			{Singh}}, \bibinfo {author} {\bibfnamefont {B.~H.}\ \bibnamefont
			{Schneider}}, \bibinfo {author} {\bibfnamefont {S.~J.}\ \bibnamefont
			{Bosman}}, \bibinfo {author} {\bibfnamefont {E.~P.~J.}\ \bibnamefont
			{Merkx}}, \ and\ \bibinfo {author} {\bibfnamefont {G.~A.}\ \bibnamefont
			{Steele}},\ }\href {\doibase 10.1063/1.4903042} {\bibfield  {journal}
		{\bibinfo  {journal} {Applied Physics Letters}\ }\textbf {\bibinfo {volume}
			{105}},\ \bibinfo {pages} {222601} (\bibinfo {year} {2014})}\BibitemShut
	{NoStop}%
	\bibitem [{\citenamefont {Kwon}\ \emph {et~al.}(2018)\citenamefont {Kwon},
		\citenamefont {{Fadavi Roudsari}}, \citenamefont {Benningshof}, \citenamefont
		{Tang}, \citenamefont {Mohebbi}, \citenamefont {Taminiau}, \citenamefont
		{Langenberg}, \citenamefont {Lee}, \citenamefont {Nichols}, \citenamefont
		{Cory},\ and\ \citenamefont {Miao}}]{Kwon2018}%
	\BibitemOpen
	\bibfield  {author} {\bibinfo {author} {\bibfnamefont {S.}~\bibnamefont
			{Kwon}}, \bibinfo {author} {\bibfnamefont {A.}~\bibnamefont {{Fadavi
					Roudsari}}}, \bibinfo {author} {\bibfnamefont {O.~W.~B.}\ \bibnamefont
			{Benningshof}}, \bibinfo {author} {\bibfnamefont {Y.~C.}\ \bibnamefont
			{Tang}}, \bibinfo {author} {\bibfnamefont {H.~R.}\ \bibnamefont {Mohebbi}},
		\bibinfo {author} {\bibfnamefont {I.~A.~J.}\ \bibnamefont {Taminiau}},
		\bibinfo {author} {\bibfnamefont {D.}~\bibnamefont {Langenberg}}, \bibinfo
		{author} {\bibfnamefont {S.}~\bibnamefont {Lee}}, \bibinfo {author}
		{\bibfnamefont {G.}~\bibnamefont {Nichols}}, \bibinfo {author} {\bibfnamefont
			{D.~G.}\ \bibnamefont {Cory}}, \ and\ \bibinfo {author} {\bibfnamefont
			{G.~X.}\ \bibnamefont {Miao}},\ }\href {\doibase 10.1063/1.5027003}
	{\bibfield  {journal} {\bibinfo  {journal} {Journal of Applied Physics}\
		}\textbf {\bibinfo {volume} {124}} (\bibinfo {year} {2018}),\
		10.1063/1.5027003}\BibitemShut {NoStop}%
	\bibitem [{\citenamefont {Ghirri}\ \emph {et~al.}(2015)\citenamefont {Ghirri},
		\citenamefont {Bonizzoni}, \citenamefont {Gerace}, \citenamefont {Sanna},
		\citenamefont {Cassinese},\ and\ \citenamefont {Affronte}}]{YBCO}%
	\BibitemOpen
	\bibfield  {author} {\bibinfo {author} {\bibfnamefont {A.}~\bibnamefont
			{Ghirri}}, \bibinfo {author} {\bibfnamefont {C.}~\bibnamefont {Bonizzoni}},
		\bibinfo {author} {\bibfnamefont {D.}~\bibnamefont {Gerace}}, \bibinfo
		{author} {\bibfnamefont {S.}~\bibnamefont {Sanna}}, \bibinfo {author}
		{\bibfnamefont {A.}~\bibnamefont {Cassinese}}, \ and\ \bibinfo {author}
		{\bibfnamefont {M.}~\bibnamefont {Affronte}},\ }\href {\doibase
		10.1063/1.4920930} {\bibfield  {journal} {\bibinfo  {journal} {Applied
				Physics Letters}\ }\textbf {\bibinfo {volume} {106}},\ \bibinfo {pages}
		{184101} (\bibinfo {year} {2015})}\BibitemShut {NoStop}%
	\bibitem [{\citenamefont {Bothner}\ \emph {et~al.}(2011)\citenamefont
		{Bothner}, \citenamefont {Gaber}, \citenamefont {Kemmler}, \citenamefont
		{Koelle},\ and\ \citenamefont {Kleiner}}]{Bothner2011}%
	\BibitemOpen
	\bibfield  {author} {\bibinfo {author} {\bibfnamefont {D.}~\bibnamefont
			{Bothner}}, \bibinfo {author} {\bibfnamefont {T.}~\bibnamefont {Gaber}},
		\bibinfo {author} {\bibfnamefont {M.}~\bibnamefont {Kemmler}}, \bibinfo
		{author} {\bibfnamefont {D.}~\bibnamefont {Koelle}}, \ and\ \bibinfo {author}
		{\bibfnamefont {R.}~\bibnamefont {Kleiner}},\ }\href {\doibase
		10.1063/1.3560480} {\bibfield  {journal} {\bibinfo  {journal} {Applied
				Physics Letters}\ }\textbf {\bibinfo {volume} {98}},\ \bibinfo {pages}
		{102504} (\bibinfo {year} {2011})}\BibitemShut {NoStop}%
	\bibitem [{\citenamefont {Martinis}(2005)}]{martinis}%
	\BibitemOpen
	\bibfield  {author} {\bibinfo {author} {\bibfnamefont {J.~M.}\ \bibnamefont
			{Martinis}},\ }\href@noop {} {\bibfield  {journal} {\bibinfo  {journal}
			{Physical Review Letters}\ }\textbf {\bibinfo {volume} {92}} (\bibinfo {year}
		{2005})}\BibitemShut {NoStop}%
	\bibitem [{\citenamefont {Kuit}\ \emph {et~al.}(2008)\citenamefont {Kuit},
		\citenamefont {Kirtley}, \citenamefont {van~der Veur}, \citenamefont
		{Molenaar}, \citenamefont {Roesthuis}, \citenamefont {Troeman}, \citenamefont
		{Clem}, \citenamefont {Hilgenkamp}, \citenamefont {Rogalla},\ and\
		\citenamefont {Flokstra}}]{kuit}%
	\BibitemOpen
	\bibfield  {author} {\bibinfo {author} {\bibfnamefont {K.~H.}\ \bibnamefont
			{Kuit}}, \bibinfo {author} {\bibfnamefont {J.~R.}\ \bibnamefont {Kirtley}},
		\bibinfo {author} {\bibfnamefont {W.}~\bibnamefont {van~der Veur}}, \bibinfo
		{author} {\bibfnamefont {C.~G.}\ \bibnamefont {Molenaar}}, \bibinfo {author}
		{\bibfnamefont {F.~J.~G.}\ \bibnamefont {Roesthuis}}, \bibinfo {author}
		{\bibfnamefont {A.~G.~P.}\ \bibnamefont {Troeman}}, \bibinfo {author}
		{\bibfnamefont {J.~R.}\ \bibnamefont {Clem}}, \bibinfo {author}
		{\bibfnamefont {H.}~\bibnamefont {Hilgenkamp}}, \bibinfo {author}
		{\bibfnamefont {H.}~\bibnamefont {Rogalla}}, \ and\ \bibinfo {author}
		{\bibfnamefont {J.}~\bibnamefont {Flokstra}},\ }\href {\doibase
		10.1103/PhysRevB.77.134504} {\bibfield  {journal} {\bibinfo  {journal}
			{Physical Review B}\ }\textbf {\bibinfo {volume} {77}},\ \bibinfo {pages}
		{134504} (\bibinfo {year} {2008})}\BibitemShut {NoStop}%
	\bibitem [{\citenamefont {de~Graaf}\ \emph {et~al.}(2013)\citenamefont
		{de~Graaf}, \citenamefont {Danilov}, \citenamefont {Adamyan},\ and\
		\citenamefont {Kubatkin}}]{degraaf}%
	\BibitemOpen
	\bibfield  {author} {\bibinfo {author} {\bibfnamefont {S.~E.}\ \bibnamefont
			{de~Graaf}}, \bibinfo {author} {\bibfnamefont {A.~V.}\ \bibnamefont
			{Danilov}}, \bibinfo {author} {\bibfnamefont {A.}~\bibnamefont {Adamyan}}, \
		and\ \bibinfo {author} {\bibfnamefont {S.~E.}\ \bibnamefont {Kubatkin}},\
	}\href {\doibase 10.1063/1.4792381} {\bibfield  {journal} {\bibinfo
			{journal} {Review of Scientific Instruments}\ }\textbf {\bibinfo {volume}
			{84}},\ \bibinfo {pages} {023706} (\bibinfo {year} {2013})}\BibitemShut
	{NoStop}%
	\bibitem [{\citenamefont {Samkharadze}\ \emph {et~al.}(2016)\citenamefont
		{Samkharadze}, \citenamefont {Bruno}, \citenamefont {Scarlino}, \citenamefont
		{Zheng}, \citenamefont {Divincenzo}, \citenamefont {Dicarlo},\ and\
		\citenamefont {Vandersypen}}]{Samkharadze2016}%
	\BibitemOpen
	\bibfield  {author} {\bibinfo {author} {\bibfnamefont {N.}~\bibnamefont
			{Samkharadze}}, \bibinfo {author} {\bibfnamefont {A.}~\bibnamefont {Bruno}},
		\bibinfo {author} {\bibfnamefont {P.}~\bibnamefont {Scarlino}}, \bibinfo
		{author} {\bibfnamefont {G.}~\bibnamefont {Zheng}}, \bibinfo {author}
		{\bibfnamefont {D.~P.}\ \bibnamefont {Divincenzo}}, \bibinfo {author}
		{\bibfnamefont {L.}~\bibnamefont {Dicarlo}}, \ and\ \bibinfo {author}
		{\bibfnamefont {L.~M.~K.}\ \bibnamefont {Vandersypen}},\ }\href {\doibase
		10.1103/PhysRevApplied.5.044004} {\bibfield  {journal} {\bibinfo  {journal}
			{Physical Review Applied}\ }\textbf {\bibinfo {volume} {5}},\ \bibinfo
		{pages} {044004} (\bibinfo {year} {2016})}\BibitemShut {NoStop}%
	\bibitem [{\citenamefont {Thoen}\ \emph {et~al.}(2017)\citenamefont {Thoen},
		\citenamefont {Bos}, \citenamefont {Haalebos}, \citenamefont {Klapwijk},
		\citenamefont {Baselmans},\ and\ \citenamefont {Endo}}]{Thoen2017}%
	\BibitemOpen
	\bibfield  {author} {\bibinfo {author} {\bibfnamefont {D.~J.}\ \bibnamefont
			{Thoen}}, \bibinfo {author} {\bibfnamefont {B.~G.~C.}\ \bibnamefont {Bos}},
		\bibinfo {author} {\bibfnamefont {E.~A.~F.}\ \bibnamefont {Haalebos}},
		\bibinfo {author} {\bibfnamefont {T.~M.}\ \bibnamefont {Klapwijk}}, \bibinfo
		{author} {\bibfnamefont {J.~J.~A.}\ \bibnamefont {Baselmans}}, \ and\
		\bibinfo {author} {\bibfnamefont {A.}~\bibnamefont {Endo}},\ }\href {\doibase
		10.1109/TASC.2016.2631948} {\bibfield  {journal} {\bibinfo  {journal} {IEEE
				Transactions on Applied Superconductivity}\ }\textbf {\bibinfo {volume}
			{27}},\ \bibinfo {pages} {1} (\bibinfo {year} {2017})}\BibitemShut {NoStop}%
	\bibitem [{\citenamefont {Nsanzineza}\ and\ \citenamefont
		{Plourde}(2014)}]{plourde}%
	\BibitemOpen
	\bibfield  {author} {\bibinfo {author} {\bibfnamefont {I.}~\bibnamefont
			{Nsanzineza}}\ and\ \bibinfo {author} {\bibfnamefont {B.~L.~T.}\ \bibnamefont
			{Plourde}},\ }\href {\doibase 10.1103/PhysRevLett.113.117002} {\bibfield
		{journal} {\bibinfo  {journal} {Physical Review Letters}\ }\textbf {\bibinfo
			{volume} {113}},\ \bibinfo {pages} {117002} (\bibinfo {year}
		{2014})}\BibitemShut {NoStop}%
	\bibitem [{\citenamefont {Lee}\ \emph {et~al.}(1999)\citenamefont {Lee},
		\citenamefont {Janko}, \citenamefont {Derenyi},\ and\ \citenamefont
		{Barabasi}}]{ratchet}%
	\BibitemOpen
	\bibfield  {author} {\bibinfo {author} {\bibfnamefont {C.~S.}\ \bibnamefont
			{Lee}}, \bibinfo {author} {\bibfnamefont {B.}~\bibnamefont {Janko}}, \bibinfo
		{author} {\bibfnamefont {I.}~\bibnamefont {Derenyi}}, \ and\ \bibinfo
		{author} {\bibfnamefont {A.~L.}\ \bibnamefont {Barabasi}},\ }\href@noop {}
	{\bibfield  {journal} {\bibinfo  {journal} {Nature}\ }\textbf {\bibinfo
			{volume} {400}},\ \bibinfo {pages} {337} (\bibinfo {year}
		{1999})}\BibitemShut {NoStop}%
	\bibitem [{\citenamefont {Bothner}\ \emph {et~al.}(2012)\citenamefont
		{Bothner}, \citenamefont {Gaber}, \citenamefont {Kemmler}, \citenamefont
		{Koelle}, \citenamefont {Kleiner}, \citenamefont {W{\"{u}}nsch},\ and\
		\citenamefont {Siegel}}]{Bothner2012}%
	\BibitemOpen
	\bibfield  {author} {\bibinfo {author} {\bibfnamefont {D.}~\bibnamefont
			{Bothner}}, \bibinfo {author} {\bibfnamefont {T.}~\bibnamefont {Gaber}},
		\bibinfo {author} {\bibfnamefont {M.}~\bibnamefont {Kemmler}}, \bibinfo
		{author} {\bibfnamefont {D.}~\bibnamefont {Koelle}}, \bibinfo {author}
		{\bibfnamefont {R.}~\bibnamefont {Kleiner}}, \bibinfo {author} {\bibfnamefont
			{S.}~\bibnamefont {W{\"{u}}nsch}}, \ and\ \bibinfo {author} {\bibfnamefont
			{M.}~\bibnamefont {Siegel}},\ }\href {\doibase 10.1103/PhysRevB.86.014517}
	{\bibfield  {journal} {\bibinfo  {journal} {Physical Review B}\ }\textbf
		{\bibinfo {volume} {86}},\ \bibinfo {pages} {014517} (\bibinfo {year}
		{2012})}\BibitemShut {NoStop}%
	\bibitem [{\citenamefont {Woods}\ \emph {et~al.}(2018)\citenamefont {Woods},
		\citenamefont {Calusine}, \citenamefont {Melville}, \citenamefont {Sevi},
		\citenamefont {Golden}, \citenamefont {Kim}, \citenamefont {Rosenberg},
		\citenamefont {Yoder},\ and\ \citenamefont {Oliver}}]{Woods2018}%
	\BibitemOpen
	\bibfield  {author} {\bibinfo {author} {\bibfnamefont {W.}~\bibnamefont
			{Woods}}, \bibinfo {author} {\bibfnamefont {G.}~\bibnamefont {Calusine}},
		\bibinfo {author} {\bibfnamefont {A.}~\bibnamefont {Melville}}, \bibinfo
		{author} {\bibfnamefont {A.}~\bibnamefont {Sevi}}, \bibinfo {author}
		{\bibfnamefont {E.}~\bibnamefont {Golden}}, \bibinfo {author} {\bibfnamefont
			{D.~K.}\ \bibnamefont {Kim}}, \bibinfo {author} {\bibfnamefont
			{D.}~\bibnamefont {Rosenberg}}, \bibinfo {author} {\bibfnamefont {J.~L.}\
			\bibnamefont {Yoder}}, \ and\ \bibinfo {author} {\bibfnamefont {W.~D.}\
			\bibnamefont {Oliver}},\ }\href@noop {} {\  (\bibinfo {year} {2018})},\
	\Eprint {http://arxiv.org/abs/1808.10347v1} {arXiv:1808.10347v1} \BibitemShut
	{NoStop}%
	\bibitem [{\citenamefont {Latimer}\ \emph {et~al.}(2012)\citenamefont
		{Latimer}, \citenamefont {Berdiyorov}, \citenamefont {Xiao}, \citenamefont
		{Kwok},\ and\ \citenamefont {Peeters}}]{interstitial1}%
	\BibitemOpen
	\bibfield  {author} {\bibinfo {author} {\bibfnamefont {M.~L.}\ \bibnamefont
			{Latimer}}, \bibinfo {author} {\bibfnamefont {G.~R.}\ \bibnamefont
			{Berdiyorov}}, \bibinfo {author} {\bibfnamefont {Z.~L.}\ \bibnamefont
			{Xiao}}, \bibinfo {author} {\bibfnamefont {W.~K.}\ \bibnamefont {Kwok}}, \
		and\ \bibinfo {author} {\bibfnamefont {F.~M.}\ \bibnamefont {Peeters}},\
	}\href {\doibase 10.1103/PhysRevB.85.012505} {\bibfield  {journal} {\bibinfo
			{journal} {Physical Review B}\ }\textbf {\bibinfo {volume} {85}},\ \bibinfo
		{pages} {12505} (\bibinfo {year} {2012})}\BibitemShut {NoStop}%
	\bibitem [{\citenamefont {Velez}\ \emph {et~al.}(2002)\citenamefont {Velez},
		\citenamefont {Jaque}, \citenamefont {Mart$\backslash$'$\backslash$in},
		\citenamefont {Montero}, \citenamefont {Schuller},\ and\ \citenamefont
		{Vicent}}]{interstitial2}%
	\BibitemOpen
	\bibfield  {author} {\bibinfo {author} {\bibfnamefont {M.}~\bibnamefont
			{Velez}}, \bibinfo {author} {\bibfnamefont {D.}~\bibnamefont {Jaque}},
		\bibinfo {author} {\bibfnamefont {J.~I.}\ \bibnamefont
			{Mart$\backslash$'$\backslash$in}}, \bibinfo {author} {\bibfnamefont {M.~I.}\
			\bibnamefont {Montero}}, \bibinfo {author} {\bibfnamefont {I.~K.}\
			\bibnamefont {Schuller}}, \ and\ \bibinfo {author} {\bibfnamefont {J.~L.}\
			\bibnamefont {Vicent}},\ }\href {\doibase 10.1103/PhysRevB.65.104511}
	{\bibfield  {journal} {\bibinfo  {journal} {Physical Review B}\ }\textbf
		{\bibinfo {volume} {65}},\ \bibinfo {pages} {104511} (\bibinfo {year}
		{2002})}\BibitemShut {NoStop}%
	\bibitem [{\citenamefont {Moshchalkov}\ \emph {et~al.}(1998)\citenamefont
		{Moshchalkov}, \citenamefont {Baert}, \citenamefont {Metlushko},
		\citenamefont {Rosseel}, \citenamefont {{Van Bael}}, \citenamefont {Temst},
		\citenamefont {Bruynseraede},\ and\ \citenamefont
		{Jonckheere}}]{interstitial3}%
	\BibitemOpen
	\bibfield  {author} {\bibinfo {author} {\bibfnamefont {V.~V.}\ \bibnamefont
			{Moshchalkov}}, \bibinfo {author} {\bibfnamefont {M.}~\bibnamefont {Baert}},
		\bibinfo {author} {\bibfnamefont {V.~V.}\ \bibnamefont {Metlushko}}, \bibinfo
		{author} {\bibfnamefont {E.}~\bibnamefont {Rosseel}}, \bibinfo {author}
		{\bibfnamefont {M.~J.}\ \bibnamefont {{Van Bael}}}, \bibinfo {author}
		{\bibfnamefont {K.}~\bibnamefont {Temst}}, \bibinfo {author} {\bibfnamefont
			{Y.}~\bibnamefont {Bruynseraede}}, \ and\ \bibinfo {author} {\bibfnamefont
			{R.}~\bibnamefont {Jonckheere}},\ }\href {\doibase 10.1103/PhysRevB.57.3615}
	{\bibfield  {journal} {\bibinfo  {journal} {Physical Review B}\ }\textbf
		{\bibinfo {volume} {57}},\ \bibinfo {pages} {3615} (\bibinfo {year}
		{1998})}\BibitemShut {NoStop}%
	\bibitem [{\citenamefont {Frey}\ \emph {et~al.}(2012)\citenamefont {Frey},
		\citenamefont {Leek}, \citenamefont {Beck}, \citenamefont {Blais},
		\citenamefont {Ihn}, \citenamefont {Ensslin},\ and\ \citenamefont
		{Wallraff}}]{Frey}%
	\BibitemOpen
	\bibfield  {author} {\bibinfo {author} {\bibfnamefont {T.}~\bibnamefont
			{Frey}}, \bibinfo {author} {\bibfnamefont {P.~J.}\ \bibnamefont {Leek}},
		\bibinfo {author} {\bibfnamefont {M.}~\bibnamefont {Beck}}, \bibinfo {author}
		{\bibfnamefont {A.}~\bibnamefont {Blais}}, \bibinfo {author} {\bibfnamefont
			{T.}~\bibnamefont {Ihn}}, \bibinfo {author} {\bibfnamefont {K.}~\bibnamefont
			{Ensslin}}, \ and\ \bibinfo {author} {\bibfnamefont {A.}~\bibnamefont
			{Wallraff}},\ }\href {\doibase 10.1103/PhysRevLett.108.046807} {\bibfield
		{journal} {\bibinfo  {journal} {Physical Review Letters}\ }\textbf {\bibinfo
			{volume} {108}} (\bibinfo {year} {2012}),\
		10.1103/PhysRevLett.108.046807}\BibitemShut {NoStop}%
	\bibitem [{\citenamefont {Ranjan}\ \emph {et~al.}(2015)\citenamefont {Ranjan},
		\citenamefont {Puebla-Hellmann}, \citenamefont {Jung}, \citenamefont
		{Hasler}, \citenamefont {Nunnenkamp}, \citenamefont {Muoth}, \citenamefont
		{Hierold}, \citenamefont {Wallraff},\ and\ \citenamefont
		{Sch{\"{o}}nenberger}}]{Ranjan2015}%
	\BibitemOpen
	\bibfield  {author} {\bibinfo {author} {\bibfnamefont {V.}~\bibnamefont
			{Ranjan}}, \bibinfo {author} {\bibfnamefont {G.}~\bibnamefont
			{Puebla-Hellmann}}, \bibinfo {author} {\bibfnamefont {M.}~\bibnamefont
			{Jung}}, \bibinfo {author} {\bibfnamefont {T.}~\bibnamefont {Hasler}},
		\bibinfo {author} {\bibfnamefont {A.}~\bibnamefont {Nunnenkamp}}, \bibinfo
		{author} {\bibfnamefont {M.}~\bibnamefont {Muoth}}, \bibinfo {author}
		{\bibfnamefont {C.}~\bibnamefont {Hierold}}, \bibinfo {author} {\bibfnamefont
			{A.}~\bibnamefont {Wallraff}}, \ and\ \bibinfo {author} {\bibfnamefont
			{C.}~\bibnamefont {Sch{\"{o}}nenberger}},\ }\href {\doibase
		10.1038/ncomms8165} {\bibfield  {journal} {\bibinfo  {journal} {Nature
				Communications}\ }\textbf {\bibinfo {volume} {6}},\ \bibinfo {pages} {7165}
		(\bibinfo {year} {2015})}\BibitemShut {NoStop}%
	\bibitem [{\citenamefont {Wang}\ \emph {et~al.}(2016)\citenamefont {Wang},
		\citenamefont {Deacon}, \citenamefont {Car}, \citenamefont {Bakkers},\ and\
		\citenamefont {Ishibashi}}]{Wang2016}%
	\BibitemOpen
	\bibfield  {author} {\bibinfo {author} {\bibfnamefont {R.}~\bibnamefont
			{Wang}}, \bibinfo {author} {\bibfnamefont {R.~S.}\ \bibnamefont {Deacon}},
		\bibinfo {author} {\bibfnamefont {D.}~\bibnamefont {Car}}, \bibinfo {author}
		{\bibfnamefont {E.~P. A.~M.}\ \bibnamefont {Bakkers}}, \ and\ \bibinfo
		{author} {\bibfnamefont {K.}~\bibnamefont {Ishibashi}},\ }\href {\doibase
		10.1063/1.4950764} {\bibfield  {journal} {\bibinfo  {journal} {Applied
				Physics Letters}\ }\textbf {\bibinfo {volume} {108}},\ \bibinfo {pages}
		{203502} (\bibinfo {year} {2016})}\BibitemShut {NoStop}%
	\bibitem [{\citenamefont {G{\"{u}}l}\ \emph {et~al.}(2017)\citenamefont
		{G{\"{u}}l}, \citenamefont {Zhang}, \citenamefont {{De Vries}}, \citenamefont
		{{Van Veen}}, \citenamefont {Zuo}, \citenamefont {Mourik}, \citenamefont
		{Conesa-Boj}, \citenamefont {Nowak}, \citenamefont {{Van Woerkom}},
		\citenamefont {Quintero-P{\'{e}}rez}, \citenamefont {Cassidy}, \citenamefont
		{Geresdi}, \citenamefont {Koelling}, \citenamefont {Car}, \citenamefont
		{Plissard}, \citenamefont {Bakkers},\ and\ \citenamefont
		{Kouwenhoven}}]{Gul2017}%
	\BibitemOpen
	\bibfield  {author} {\bibinfo {author} {\bibfnamefont {{\"{O}}.}~\bibnamefont
			{G{\"{u}}l}}, \bibinfo {author} {\bibfnamefont {H.}~\bibnamefont {Zhang}},
		\bibinfo {author} {\bibfnamefont {F.~K.}\ \bibnamefont {{De Vries}}},
		\bibinfo {author} {\bibfnamefont {J.}~\bibnamefont {{Van Veen}}}, \bibinfo
		{author} {\bibfnamefont {K.}~\bibnamefont {Zuo}}, \bibinfo {author}
		{\bibfnamefont {V.}~\bibnamefont {Mourik}}, \bibinfo {author} {\bibfnamefont
			{S.}~\bibnamefont {Conesa-Boj}}, \bibinfo {author} {\bibfnamefont {M.~P.}\
			\bibnamefont {Nowak}}, \bibinfo {author} {\bibfnamefont {D.~J.}\ \bibnamefont
			{{Van Woerkom}}}, \bibinfo {author} {\bibfnamefont {M.}~\bibnamefont
			{Quintero-P{\'{e}}rez}}, \bibinfo {author} {\bibfnamefont {M.~C.}\
			\bibnamefont {Cassidy}}, \bibinfo {author} {\bibfnamefont {A.}~\bibnamefont
			{Geresdi}}, \bibinfo {author} {\bibfnamefont {S.}~\bibnamefont {Koelling}},
		\bibinfo {author} {\bibfnamefont {D.}~\bibnamefont {Car}}, \bibinfo {author}
		{\bibfnamefont {S.~R.}\ \bibnamefont {Plissard}}, \bibinfo {author}
		{\bibfnamefont {E.~P. A.~M.}\ \bibnamefont {Bakkers}}, \ and\ \bibinfo
		{author} {\bibfnamefont {L.~P.}\ \bibnamefont {Kouwenhoven}},\ }\href
	{\doibase 10.1021/acs.nanolett.7b00540} {\bibfield  {journal} {\bibinfo
			{journal} {Nano Letters}\ }\textbf {\bibinfo {volume} {17}},\ \bibinfo
		{pages} {2690} (\bibinfo {year} {2017})}\BibitemShut {NoStop}%
\end{thebibliography}
\end{document}